\documentclass[prd,aps,preprintnumbers,fleqn,showpacs,nofootinbib,superscriptaddress]{revtex4}

\usepackage{amssymb,amsmath,graphicx,bm}

\DeclareMathOperator{\tr}{Tr}

\newcommand{\xB}{x_{\scriptscriptstyle B}}
\newcommand{\sT}{{\scriptscriptstyle T}}
\renewcommand{\d}{\mathrm{d}}
\def\slash#1{\setbox0=\hbox{$#1$}               
        \dimen0=\wd0                            
        \setbox1=\hbox{/} \dimen1=\wd1          
        \ifdim\dimen0>\dimen1                   
        \rlap{\hbox to \dimen0{\hfil/\hfil}}    
        #1                                      
        \else              
        \rlap{\hbox to \dimen1{\hfil$#1$\hfil}} 
        /                                       
        \fi}                                    %

\begin{document}

\preprint{NIKHEF 2013-023}
\preprint{SLAC-PUB-15688}

\title{Linear polarization of gluons and photons in unpolarized collider 
experiments}

\author{Cristian Pisano}
\email{c.pisano@nikhef.nl}
\affiliation{Nikhef and Department of Physics and Astronomy, 
VU University Amsterdam, De Boelelaan 1081, NL-1081 HV Amsterdam, 
The Netherlands}

\author{Dani\"el Boer}
\email{d.boer@rug.nl}
\affiliation{Theory Group, KVI, University of Groningen, 
Zernikelaan 25, NL-9747 AA Groningen, The Netherlands}

\author{Stanley J.~Brodsky}
\email{sjbth@slac.stanford.edu}
\affiliation{SLAC National Accelerator Laboratory, Stanford University, 
Stanford, California 94309, USA}

\author{Maarten G. A. Buffing}
\email{m.g.a.buffing@vu.nl}
\affiliation{Nikhef and Department of Physics and Astronomy, 
VU University Amsterdam, De Boelelaan 1081, NL-1081 HV Amsterdam, 
The Netherlands}

\author{Piet J. Mulders}
\email{mulders@few.vu.nl}
\affiliation{Nikhef and Department of Physics and Astronomy, VU University
 Amsterdam, De Boelelaan 1081, NL-1081 HV Amsterdam, 
The Netherlands}

\begin{abstract}
We study azimuthal asymmetries in heavy quark pair production in unpolarized electron-proton and proton-proton collisions, where the asymmetries originate from
the linear polarization of gluons inside unpolarized hadrons. We provide cross section expressions and study the maximal asymmetries allowed by positivity, for both charm and bottom quark pair production. The upper bounds on the asymmetries are shown to be very large depending on the transverse momentum of the heavy quarks, which is promising especially for their measurements at a possible future Electron-Ion Collider or a Large Hadron electron Collider. We also study the analogous processes and asymmetries in muon pair production as a means to probe linearly polarized photons inside unpolarized protons. For increasing invariant mass of the muon pair the asymmetries become very similar to the heavy quark pair ones. Finally, we discuss the process dependence of the results that arises due to differences in color flow and address the problem with factorization in case of proton-proton collisions.  
\end{abstract}

\date{\today}

\maketitle

\section{Introduction}
It is well-known that photons radiated off from electrons can carry linear polarization. In terms of photon helicity, it corresponds to 
an interference between $+1$ and $-1$ helicity states. Formally one can view this momentum dependent photon distribution as the distribution 
of linearly polarized photons `inside' an electron. Similarly, there are distributions of linearly polarized photons and gluons inside a proton, 
here denoted by $h_1^{\perp \, \gamma}$ and $h_1^{\perp \, g}$, respectively. The latter has received growing attention recently, because it affects 
high energy collisions involving unpolarized protons, such as at LHC. 

The distribution of linearly polarized gluons inside an unpolarized hadron was first considered in Ref.~\cite{Mulders:2000sh} and later
discussed in a model context in Ref.~\cite{Meissner:2007rx}. In Ref.~\cite{Boer:2009nc} it was noted that it contributes to the dijet imbalance in 
unpolarized hadronic collisions, which is commonly used to determine the 
average transverse momentum squared ($\langle p_T^2 \rangle$) of partons inside
 protons. Depending 
on the size of $h_1^{\perp \, g}$ and on whether its contribution can be calculated and taken into account, it may complicate or even hamper the 
determination of the average transverse momentum of partons. It is therefore important to determine its size separately using other observables. 
Although in Ref.~\cite{Boer:2009nc} it was discussed how to isolate the contribution from $h_1^{\perp \, g}$ by means of an azimuthal angular dependent 
weighting of the cross section, proton-proton collisions are expected to suffer from contributions that break factorization, through initial 
and final state interactions \cite{Rogers:2010dm}. In Ref.~\cite{Boer:2010zf} a theoretically cleaner and safer way was considered: heavy quark pair 
production in electron-proton collisions, for instance at a future 
Electron-Ion Collider. Another process, where the problem 
of factorization breaking is absent, is $pp \to \gamma \gamma X$, which was investigated in Ref.~\cite{Qiu:2011ai} specifically for RHIC. 
 
Linearly polarized gluons in proton-nucleus scattering have been considered in Refs.~\cite{Metz:2011wb,Dominguez:2011br,Schafer:2012yx,Akcakaya:2012si}, 
where factorization may also work out in the dilute-dense 
regime as discussed in Ref.~\cite{Akcakaya:2012si}. These studies also suggest that at small $x$-fractions 
of the gluons inside a nucleus, the distribution of linear polarization may reach its maximally allowed size, which is bounded by the distribution of unpolarized gluons~\cite{Mulders:2000sh}. Moreover, just like in the case of linearly polarized photons, which are perturbatively generated from electrons, linearly polarized gluons are also perturbatively generated from unpolarized quarks and gluons inside the proton 
\cite{Nadolsky:2007ba,Catani:2010pd,deFlorian:2012mx}. This determines the large transverse momentum tail of the distribution \cite{Sun:2011iw}. It shows that the tail falls off with the same power as the unpolarized gluon distribution. Therefore, the degree of polarization does not fall off with increasing transverse momentum, see figure 2 of Ref.\ \cite{Boer:2013fca}. This means that although the magnitude has not yet been determined from experiment, the expectation is that it is not small at high energy. 

It has also recently been noted that $h_1^{\perp \, g}$ affects the angular independent transverse momentum distribution of scalar or pseudoscalar 
particles, such as the Higgs boson \cite{Sun:2011iw,Boer:2011kf,Boer:2013fca} or charmonium and bottomonium states \cite{Boer:2012bt}. These results could 
help pinpoint the quantum numbers of the boson recently discovered at LHC or allow a determination of $h_1^{\perp \, g}$ at LHCb for example. These ideas are very similar to the suggestions to use linear polarization at photon colliders to investigate Higgs production 
\cite{Grzadkowski:1992sa,Gunion:1994wy,Kramer:1993jn,Gounaris:1997ef,Asner:2001ia} and heavy quark production 
\cite{Kamal:1995ct, Jikia:1996bi,Melles:1998gu,Jikia:2000rk,Kniehl:2009kh}, 
that are of interest for investigations at ILC. 
There are some notable differences with the proton distributions though: 
the transverse momentum of photons radiated off from electrons is known exactly in the photon-photon scattering case, whereas in proton-proton collisions 
the gluonic transverse momentum distributions enter in a convolution integral. Moreover, the QED case does not have the problems with non-factorizing initial and final state interactions (ISI/FSI) that arise in the non-Abelian case for certain processes. Like any other transverse momentum dependent parton distribution, the function $h_1^{\perp\, g}$ will receive contributions from ISI or FSI and is therefore 
expected to be process dependent. Apart from the fact that $h_1^{\perp\, g}$ can thus be nonuniversal, the ISI/FSI can even lead to violations of 
pQCD factorization at leading twist, as already mentioned above. By considering several different extractions, the nonuniversality and the 
factorization breaking can be studied and quantified. From this point of view it is also very interesting to compare to the linearly polarized photon 
distribution inside the proton.

In the present paper we will consider heavy quark pair production in electron-proton and proton-proton collisions. It is partly intended to provide the calculational details of the results in Ref.~\cite{Boer:2010zf}, but it also contains additional results, for example on process dependence.  Also, we include the analogues in muon pair production as a means to probe 
$h_1^{\perp \, \gamma}$, which describes linearly polarized photons inside unpolarized protons. 
The paper is organized as follows. First we will discuss electron-hadron scattering for three cases: heavy quark pair production, dijet production and 
muon pair production. After presenting the general expressions, we identify the most promising azimuthal asymmetries that will allow access to 
$h_1^{\perp \, g}$ and $h_1^{\perp \, \gamma}$. We present upper bounds on these asymmetries, for charm and bottom quarks and muons. This will hopefully 
expedite future experimental investigations of these distributions. Next we turn to hadron-hadron collisions and discuss in more detail the color flow dependence and factorization breaking issues, according to the latest insights \cite{Buffing:2011mj,Buffing:2012sz,Buffing:2013notpublishedyet}.

\section{The TMDs for unpolarized hadrons}

The information on linearly polarized gluons is encoded in the transverse 
momentum dependent correlator, for which, in this paper, we only consider 
unpolarized hadrons.
The parton correlators describe
the hadron $\rightarrow$ parton transitions and 
are defined as matrix elements on the light-front LF 
($\lambda{\cdot}n\,{\equiv}\,0$, where $n$ is a light-like vector, $n^2 = 0$,
conjugate to $P$). The correlators are parameterized in terms 
of transverse momentum dependent distribution functions (TMDs).
Specifically, at leading twist and omitting gauge links, 
the quark correlator is given by \cite{Boer:1997nt}
\begin{eqnarray}
\label{QuarkCorr}
\Phi_q(x{,} \bm p_\sT)
& = & {\int}\frac{\d(\lambda{\cdot}P)\,d^2\lambda_\sT}{(2\pi)^3}\ e^{ip\cdot\lambda}\,
\langle P |\,\overline\psi(0)\,
\psi(\lambda)\,|P\rangle\,\big\rfloor_{\text{LF}} \nonumber \\
& = & \frac{1}{2}\,
\bigg \{\,f_1^q(x{,}\bm{p}_\sT^2)\;\slash P
+i h_1^{\perp\,q}(x{,}\bm{p}_\sT^2)\;\frac{[\slash p_\sT , 
\slash P]}{2 M}\bigg \}\, , 
\end{eqnarray}
with $f_1^q(x, \bm{p}_\sT^2)$ denoting the transverse momentum dependent 
distribution of unpolarized quarks inside an unpolarized hadron, and where we have used the naming convention of Ref.~\cite{Bacchetta:2006tn}.
Its integration over $\bm{p}_\sT$ provides the well-known
light-cone momentum distribution $f_1^q(x) = q(x)$. 
The function $h_1^{\perp q}(x, \bm{p}_\sT^2)$, nowadays commonly referred to as Boer-Mulders function, is time-reversal ($T$) odd and 
can be interpreted as the distribution of transversely polarized quarks 
inside an unpolarized hadron \cite{Boer:1997nt}. It gives rise to the $\cos 2 \phi$ double Boer-Mulders asymmetry in the Drell-Yan process  
and to a violation of the Lam-Tung relation \cite{Boer:1999mm,Boer:2002ju}. 
Similarly, for an antiquark,
\begin{eqnarray}
\label{AquarkCorr}
\bar\Phi_q(x{,} \bm p_\sT)
& = & -{\int}\frac{\d(\lambda{\cdot}P)\,d^2\lambda_\sT}{(2\pi)^3}\ e^{-ip\cdot\lambda}\,
\langle P |\,\overline\psi(0)\,
\psi(\lambda)\,|P\rangle\,\big\rfloor_{\text{LF}} \nonumber \\
& = & \frac{1}{2}\,
\bigg \{\,f_1^{\bar q}(x{,}\bm{p}_\sT^2)\;\slash P
+i h_1^{\perp\,\bar q}(x{,}\bm{p}_\sT^2)\;\frac{[\slash p_\sT , 
\slash P]}{2 M}\bigg \}\, . 
\end{eqnarray}
Omitting gauge links, the gluon correlator is defined as \cite{Mulders:2000sh}
\begin{eqnarray}
\label{GluonCorr}
\Phi_g^{\mu\nu}(x{,}\bm p_\sT )
& =& \frac{n_\rho\,n_\sigma}{(p{\cdot}n)^2}
{\int}
\frac{\d(\lambda{\cdot}P)\,d^2\lambda_\sT}{(2\pi)^3}\ e^{ip\cdot\lambda}\,
\langle P|\,\tr\big[\,F^{\rho\mu}(0)\,
F^{\sigma\nu}(\lambda)\,\big]
\,|P \rangle\,\big\rfloor_{\text{LF}} \nonumber \\
&=&\frac{1}{2x}\,\bigg \{-g_\sT^{\mu\nu}\,f_1^g(x{,}\bm{p}_\sT^2)
+\bigg(\frac{p_\sT^\mu p_\sT^\nu}{M^2}\,
{+}\,g_\sT^{\mu\nu}\frac{\bm p_\sT^2}{2M^2}\bigg)
\;h_1^{\perp\,g}(x{,}\bm{p}_\sT^2) \bigg \}\, ,
\label{eq:gcorr}
\end{eqnarray}
where $F^{\mu\nu}(x)$ is the gluon field strength and $g^{\mu\nu}_{\sT}$ a 
transverse tensor given by
\begin{equation}
g^{\mu\nu}_{\sT} = 
g^{\mu\nu} - P^{\mu}n^{\nu}/P\cdot n-n^{\mu}P^{\nu}/P\cdot n\,,
\end{equation}
and where we have used the naming convention of Ref.~\cite{Meissner:2007rx}.
The transverse momentum dependent function $f_1^g(x{,}\bm{p}_\sT^2)$ describes
the distribution of unpolarized gluons inside an unpolarized hadron, and, 
integrated over ${\bm{p}}_\sT$, 
gives the familiar light-cone momentum distribution $f_1^g(x) = g(x)$. 
The function $h_1^{\perp\,g}(x{,}\bm{p}_\sT^2)$ is $T$-even 
and represents the distribution of linearly polarized gluons inside 
an unpolarized hadron.

\section{\label{sec:cs} Electron-hadron collisions: calculation of the 
cross sections} 

\subsection{Heavy quark pair production}

We consider the process 
\begin{equation}
e (\ell){+}h(P)\to e(\ell') {+} Q(K_1) {+} \bar{Q}(K_2){+}X \, ,
\end{equation}
where the four-momenta of the particles are given within brackets, and
the quark-antiquark pair is almost back-to-back in the 
plane orthogonal to the direction of the hadron and the exchanged photon. 
Following Refs.~\cite{Boer:2007nd,Boer:2009nc}, we will instead of collinear 
factorization
consider a generalized factorization scheme taking into account 
partonic transverse momenta. We make a decomposition of the momenta
where $q\equiv \ell -\ell'$ and $P$ determine the light-like directions,
\begin{equation}
P = n_+ + \frac{M^2}{2}\,n_- \approx n_+
\quad \mbox{and} \quad
q = -\xB\,n_+ + \frac{Q^2}{2\,\xB}\,n_- \approx -\xB\,P + (P\cdot q)\,n_-,
\label{eq:qexpansion}
\end{equation}
where $Q^2 = -q^2$ and $\xB = Q^2/2P\cdot q$ (up to target mass corrections).
We will thus expand in $n_+ = P$ and $n_- = n = (q+\xB\,P)/P\cdot q$. 
We note that the leptonic momenta define a plane transverse
with respect to $q$ and $P$. 
Explicitly the leptonic momenta are given by
\begin{eqnarray}
\ell & = &\frac{1-y}{y}\,\xB\,P + \frac{1}{y}\,\frac{Q^2}{2\xB}\,n
+ \frac{\sqrt{1-y}}{y}\,Q\,\hat\ell_\perp
= \frac{1-y}{y}\,\xB\,P + \frac{s}{2}\,n
+ \frac{\sqrt{1-y}}{y}\,Q\,\hat\ell_\perp, 
\\
\ell^\prime & = & \frac{1}{y}\,\xB\,P + \frac{1-y}{y}\,\frac{Q^2}{2\xB}\,n
+ \frac{\sqrt{1-y}}{y}\,Q\,\hat\ell_\perp
= \frac{1}{y}\,\xB\,P + (1-y)\,\frac{s}{2}\,n
+ \frac{\sqrt{1-y}}{y}\,Q\,\hat\ell_\perp,
\end{eqnarray}
where $y = P\cdot q/P\cdot \ell$.
The total invariant mass squared is
$s = (\ell + P)^2 = 2\,\ell\cdot P = 2\,P\cdot q/y = Q^2/\xB y$. 
The invariant mass squared of the virtual photon-target
system is given by $W^2 =(q+P)^2 = Q^2(1-\xB)/\xB$. We then have
$Q^2 = \xB ys$ and $W^2 = (1-\xB)ys$.
We expand the parton momentum using the Sudakov decomposition,
\begin{equation}
p =x\,P + p_\sT + (p\cdot P-x\,M^2)\,n \approx x\,P + p_\sT,
\label{PartonDecompositions}
\end{equation}
where $x = p\cdot n$. 
We can expand the heavy quark momenta as
\begin{eqnarray}
K_1 
&=& z_1\,(P\cdot q)\,n + \frac{M_Q^2 + \bm K_{1\perp}^2}{2z_1\,P\cdot q}\,P + K_{1\perp},
\label{eq:jetmom1}
\\
K_2 
&=& z_2\,(P\cdot q)\,n +\frac{M_Q^2 + \bm K_{2\perp}^2}{2z_2\,P\cdot q}\,P + K_{2\perp},
\end{eqnarray}
with $K_{i\perp}^2 = -\bm K_{i\perp}^2$. We denote the heavy (anti)quark 
mass with $M_Q$. For the partonic subprocess we have $p+q=K_1+K_2$, implying
$z_1+z_2 = 1$. For our discussions, we introduce the sum and difference
of the transverse heavy quark momenta, 
$K_\perp = (K_{1\perp} - K_{2\perp})/2$ and 
$q_\sT = K_{1\perp} + K_{2\perp}$ with 
$\vert q_\sT\vert \ll \vert K_\perp\vert$. In that situation,
we can use the approximate transverse momenta
$K_{1\perp} \approx K_{\perp}$ and $K_{2\perp} \approx -K_{\perp}$
denoting $M_{i\perp}^2 \approx M_\perp^2 = M_Q^2 + \bm K_\perp^2$. 
We use the Mandelstam variables 
\begin{eqnarray}
\hat s &=& (q + p)^2 
= \frac{x-\xB}{\xB}\,Q^2 = xy\,s - Q^2
= (K_1+K_2)^2 
= \frac{M_{1\perp}^2}{z_1} + \frac{M_{2\perp}^2}{z_2} 
\approx \frac{M_{\perp}^2}{z_1\,z_2} ,
\\
\hat t &=& (q-K_1)^2 = M_Q^2 - \frac{M_{1\perp}^2}{z_1} - (1-z_1)\,Q^2
\approx M_Q^2 - z_2\,(\hat s + Q^2),
\\
\hat u &=& (q-K_2)^2 = M_Q^2 - \frac{M_{2\perp}^2}{z_2} - (1-z_2)\,Q^2 
\approx M_Q^2 - z_1\,(\hat s + Q^2) ,
\end{eqnarray}
from which we obtain momentum fractions,
\begin{eqnarray}
&&
x = \xB\,\frac{\hat s + Q^2}{Q^2} = \frac{\hat s + Q^2}{y\,s}
= \xB + \frac{M_\perp^2}{y\,z_1\,z_2\,s},
\label{eq:xresult}
\\
&&
z = z_2 = \frac{1}{e^{y_1-y_2} + 1} = -\frac{\hat t - M_Q^2}{\hat s+Q^2}
\quad \mbox{and} \quad
1-z = z_1 = \frac{1}{e^{y_2-y_1} + 1} = -\frac{\hat u - M_Q^2}{\hat s+Q^2},
\label{Yexpression}
\end{eqnarray}
where we have also introduced the rapidities $y_i$ for the heavy quark momenta
(along the photon-target direction).

In analogy to Refs.~\cite{Boer:2007nd} and \cite{Boer:2009nc} 
we assume that at sufficiently high energies the cross section 
factorizes in a leptonic tensor, a soft parton correlator for the incoming 
hadron and a hard part:
\begin{eqnarray}
\d\sigma
& = &\frac{1}{2 s}\,\frac{\d^3 \ell'}{(2\pi)^3\,2 E_e^{\prime}} \frac{\d^3 K_1}{(2\pi)^3\,2 E_1}
\frac{\d^3K_2}{(2\pi)^3\,2E_2}
{\int}\d x\, \d^2\bm p_{\sT}\,(2\pi)^4
\delta^4(q {+} p {-} K_{1} {-} K_{2})
 \nonumber \\
&&\qquad \qquad\qquad \qquad\qquad\times \sum_{a, b, c}\ \frac{1}{Q^4}\, 
L(\ell,q)\otimes\Phi_a(x{,}\bm p_{\sT})
\otimes\, |H_{\gamma^*\, a \rightarrow b\, c }(q, p, K_{1}, K_2)|^2\, , 
\label{CrossSec}
\end{eqnarray}
where the leptonic tensor $L(\ell, q) $ is given by
\begin{equation}
L^{\mu\nu}(\ell, q) = -g^{\mu\nu}\,Q^2 + 2\,( 
 \ell^{\mu}\ell'^\nu + \ell^{\nu}\ell'^\mu).
\end{equation}
In Eq.~(\ref{CrossSec}) the sum runs over all the partons in the initial and 
final states, and $H_{\gamma^*a\to bc}$ is the amplitude for 
the hard partonic subprocess
$\gamma^*a\to bc$. The convolutions $\otimes$ denote appropriate traces 
over the Dirac indices.

In order to derive an expression for the cross section in terms of parton 
distributions, we insert the parametrizations in Eqs.~(\ref{QuarkCorr}), 
(\ref{AquarkCorr}) and (\ref{GluonCorr}) of the TMD 
correlators into Eq.~(\ref{CrossSec}). In a frame where the 
virtual photon and the incoming hadron move along the $z$ axis, and the 
lepton scattering plane defines the azimuthal angle $\phi_\ell =\phi_{\ell^\prime}=0$, one has
\begin{equation}
\frac{\d^3 \ell'}{(2\pi)^3\, 2 E_e^{\prime}} 
= \frac{1}{16 \pi^2}\, {s}{y} \,\d\xB\,\d y\, ,
\quad \mbox{and} \quad 
\d y_i = \frac{\d z_i}{z_1 z_2}~.
\end{equation}
With the 
decompositions of the parton momenta in Eq.\ \eqref{PartonDecompositions}, 
the $\delta$-function in Eq.~(\ref{CrossSec}) can be rewritten as
\begin{equation}
\label{DeltaFunc}
\delta^4(p+q-K_1-K_2)
=\delta\bigg(x-\xB -\frac{M_\perp^2}{yz_1z_2\,s}\bigg)
\,\delta\bigg(\frac{ys}{2}(1-z_1-z_2)\bigg)
\,\delta^2\bigg(\bm p_\sT-\bm q_\sT\bigg),
\end{equation}
with corrections of order $\mathcal O(1/s)$. 
After integration over $x$ and $\bm p_\sT$, one obtains from the first
and last $\delta$-functions on the r.h.s.\ of Eq.~(\ref{DeltaFunc}), 
relations of $x$ in terms of other kinematical variables
[Eq.~(\ref{eq:xresult})], while $\bm p_\sT$ is related to the sum of the
transverse momenta of the heavy quarks, $\bm p_\sT = \bm q_\sT$.
Hence the complete angular structure of the cross section is as follows:
\begin{eqnarray}
\frac{\d\sigma}
{\d y_1\,\d y_2\,\d y\,\d\xB\,\d^2\bm{q}_{\sT} \d^2\bm{K}_{\perp}} & = &
\frac{\alpha^2\alpha_s}{\pi s M_\perp^2}\, \frac{1}{\xB y^2}\, 
\bigg\{ A_0 + A_{1}\cos \phi_\perp + A_{2} \cos 2 \phi_\perp + \bm q_\sT^2 \, \left [B_0 \cos 2 (\phi_\perp-\phi_\sT)
\right . \nonumber \\
&& \qquad +\, B_{1} \cos (\phi_\perp-2\phi_\sT) + B^{\prime}_{1} \cos (3 \phi_\perp-2\phi_\sT) + B_{2} \cos 2 \phi_\sT \nonumber \\
&& \qquad\qquad \qquad + \,\left . B^{\prime}_{2} \cos 2(2\phi_\perp-\phi_\sT) \right ] \bigg\}\,\delta(1-z_1-z_2)\,,
\label{eq:cscomplete}
\end{eqnarray}
with $\phi_\sT$ and $\phi_\perp$ denoting 
the azimuthal angles of $\bm{q}_\sT$ and $\bm{K}_\perp$,
respectively. The terms $A_i$, $B_i$, with $i=0,1,2$, and $B_{1,2}^\prime$, calculated at leading order (LO) in perturbative QCD,
are given explicitly in the following. For this calculation we
 have used the approximations discussed
above, which are applicable in the situation in which the outgoing heavy
quark and antiquark are almost back to back in the transverse plane, 
implying
$|\bm{q}_\sT| \ll |\bm{K}_{\perp}| $. In order to access experimentally 
$A_1$, $B_1$ and $B_1^\prime$, the measurement of the electric 
charge of both the heavy quark and antiquark is required. 
This would allow one to distinguish between the two of them, 
avoiding the $\cos\phi_\perp$, 
$\cos (\phi_\perp-2\phi_\sT)$ and $\cos (3 \phi_\perp-2\phi_\sT)$
 modulations from averaging out \cite{Mirkes:1997ru}. 
The terms $A_i$ in Eq.~(\ref{eq:cso}) are given by the sum 
of several contributions ${\cal A}_i^{e a\to e b c}$ coming from the 
partonic subprocesses $e a\to e b c$ underlying the reaction 
$e h\to e Q \bar{Q} X$,
\begin{equation}
A_i^{e h\to e {Q} \bar{Q} X}
 = e_Q^2 \,T_R\, {\cal A}_i^{e g\to e Q \bar Q} \, f_1^g (x,\bm{q}_{\sT}^2)\, ,\qquad i =0,1,2\,,
\label{eq:AQQb}
\end{equation}
with $T_R=1/2$. They obey the relations
\begin{eqnarray}
{\cal A}_0^{e g\to e Q \bar Q} & = & [1+(1-y)^2]\,{\cal A}_{U+L}^{\gamma^* g\to Q \bar Q} \,
-y^2 \,{\cal A}_{L}^{\gamma^* g\to Q \bar Q}\,, \nonumber \\
{\cal A}_1^{e g\to e Q \bar Q} & = & (2-y) \sqrt{1-y}\,{\cal A}_{I}^{\gamma^* g\to Q \bar Q}\,,\nonumber \\
 {\cal A}_2^{e g\to e Q \bar Q} & = & 2 (1-y)\, {\cal A}_{T}^{\gamma^* g\to Q \bar Q}\,,
\label{eq:Agstar} 
\end{eqnarray} 
where we have introduced the following linear combinations of helicity amplitudes squared ${\cal A}_{\lambda_\gamma,\lambda_\gamma^\prime}$ for the process $\gamma^* g \to Q\bar Q$ ($\lambda_\gamma, \lambda_\gamma^\prime =0, \pm 1$) \cite{Brodkorb:1994de}:
\begin{eqnarray}
{\cal A}_{U+L} & \sim & {\cal A}_{++} + {\cal A}_{--} + {\cal A}_{00}\, ,
\nonumber \\
{\cal A}_{L} & \sim & {\cal A}_{00}\,, \nonumber \\
{\cal A}_{I} & \sim & {\cal A}_{0+} +{\cal A}_{+0}-{\cal A}_{0-} -{\cal A}_{-0} \,, \nonumber \\
{\cal A}_{T} & \sim & {\cal A}_{+-} + {\cal A}_{-+}~.
\label{eq:Ahel}
\end{eqnarray}
 We find 
\begin{eqnarray}
{\cal A}_{U+L}^{\gamma^* g\to Q \bar Q} & = & \frac{1}{D^3} -\frac{z(1-z)}{D^3}\,\left \{ 2 -4\,\frac{M_Q^2}{M_\perp^2} + 4\,\frac{M_Q^4}{M_\perp^4} - \left [4z(1-z)\left (2 -3\,\frac{M_Q^2}{M_\perp^2}\right ) +2\,\frac{M_Q^2}{M_\perp^2} \right ]\frac{Q^2}{M_\perp^2} \,\right . \nonumber \\
& & \qquad\qquad \left . -\, z(1-z)[1-2 z(1-z)]\frac{Q^4}{M_\perp^2} \right \}\,, \label{eq:AQQb2}\\
{\cal A}_{L}^{\gamma^* g\to Q \bar Q} & = & 8\, \frac{z^2(1-z)^2}{D^3} \left (1-\frac{M_Q^2}{M_\perp^2}\right )\frac{Q^2}{M_\perp^2}\,, \\
{\cal A}_{I}^{\gamma^* g\to Q \bar Q} & = & 4 \,\sqrt{1-\frac{M_Q^2}{M_\perp^2}} \,
\frac{z(1-z)(1-2 z)}{D^3}\,\frac{Q}{M_\perp}
\bigg [1- z (1-z) \frac{Q^2}{M_\perp^2} -2 \frac{M_Q^2}{M_\perp^2} \bigg ]\,,
\label{eq:AQQbphil}\\
{\cal A}_{T}^{\gamma^* g\to Q \bar Q} & = & 4\,\frac{z(1-z)}{D^3}\,\bigg (1-\frac{M_Q^2}{M_\perp^2} \bigg) 
\bigg [ z (1-z) \frac{Q^2}{M_\perp^2} + \frac{M_Q^2}{M_\perp^2} \bigg ]\, ,
\label{eq:AQQb2phil}
\end{eqnarray}
where the denominator $D$ is defined as
\begin{equation}
D \equiv D \left (z,\frac{Q^2}{M_\perp^2} \right ) = 1 + z (1-z) \frac{Q^2}{M_\perp^2}~.
\label{eq:Den} 
\end{equation}
The remaining terms in Eq.~(\ref{eq:cscomplete}) depend on the polarized 
gluon distribution $h_1^{\perp\,g}(x, \bm q_\sT^2)$ and have the following 
general form,
\begin{eqnarray}
B_i^{e h\to e {Q} \bar{Q} X} & = &
\frac{1}{M^2}\, e^2_Q\,T_R\,{\cal B}_{i}^{e g\to e Q \bar Q} \, h_1^{\perp \,g} (x, \bm{q}_{\sT}^2)\, ,\qquad i = 0,1,2, \nonumber \\
B_{1, 2}^{\prime \,e h\to e {Q} \bar{Q} X} & = &
\frac{1}{M^2}\, e^2_Q\,T_R\, {\cal B}_{1, 2}^{\prime \,e g\to e Q \bar Q} \, h_1^{\perp \,g} (x, \bm{q}_{\sT}^2)\, ,
\label{eq:BQQbphi} 
\end{eqnarray}
where, in analogy to Eq.\ (\ref{eq:Agstar}), one can write
\begin{eqnarray}
{\cal B}_0^{e g\to e Q \bar Q} & = & [1+(1-y)^2]\,{\cal B}_{U+L}^{\gamma^* g\to Q \bar Q} 
-y^2 \,{\cal B}_{L}^{\gamma^* g\to Q \bar Q}\,, \nonumber \\
{\cal B}_1^{e g\to e Q \bar Q} & = & (2-y) \sqrt{1-y}\,{\cal B}_{I}^{\gamma^* g\to Q \bar Q}\,, \hspace{2.2cm} {\cal B}_1^{\prime \,e g\to e Q \bar Q} \, = \, (2-y) \sqrt{1-y}\,{\cal B}_{I}^{\prime\,\gamma^* g\to Q \bar Q}\,, \nonumber \\
 {\cal B}_2^{e g\to e Q \bar Q} & = & 2 (1-y)\, {\cal B}_{T}^{\gamma^* g\to Q \bar Q}\,,\hspace{3.18cm} {\cal B}_2^{\prime\, e g\to e Q \bar Q} \, = \, 2 (1-y)\, {\cal B}_{T}^{\prime\, \gamma^* g\to Q \bar Q}\,,
\label{eq:Bgstar} 
\end{eqnarray}
with
\begin{eqnarray}
{\cal B}_{U+L}^{e g\to e Q\bar Q}& = & \frac{z (1-z)}{D^3}\, \left (1-\frac{M_Q^2}{M_\perp^2} \right ) \left \{ \left[-1 + 6 z (1-z) \right] \frac{Q^2}{M_\perp^2} +2 \,\frac{M_Q^2}{M_\perp^2}\right\}\,,
\label{eq:B0QQb} \\
{\cal B}_L^{\gamma^* g\to Q\bar Q}& = & 4\, \frac{z^2 (1-z)^2}{D^3} \left (1-\frac{M_Q^2}{M_\perp^2} \right )\frac{Q^2}{M_\perp^2}\,, \label{eq:B0LQQb} \,\\
{\cal B}_{I}^{\gamma^* g\to Q\bar Q} & = & - 2 \,\sqrt{1-\frac{M_Q^2}{M_\perp^2}} \,
\frac{z(1-z)(1-2 z)}{D^3}\,\frac{Q}{M_\perp}
\bigg [z (1-z) \frac{Q^2}{M_\perp^2} + \frac{M_Q^2}{M_\perp^2} \bigg ]\,,\\
{\cal B}_{I}^{\prime\,\gamma^* g\to Q\bar Q} & =& 2\,\bigg ( 1-\frac{M_Q^2}{M_\perp^2}\bigg)^{\frac{3}{2}} \,
\frac{z(1-z)(1-2 z)}{D^3}\, \frac{Q}{M_\perp}\,,
\label{eq:BQQb2phi} \\
{\cal B}_{T}^{e g\to e Q\bar Q} & = &- 
\frac{z(1-z)}{D^3}\, \bigg [ z (1-z)\,\frac{Q^2}{M_\perp^2} +\frac{M_Q^2}{M_\perp^2} \bigg ]^2 \, ,\\
{\cal B}_{T}^{\prime\, \gamma^* g\to Q\bar Q} & = & -
\frac{z(1-z)}{D^3}\, \bigg ( 1-\frac{M_Q^2}{M_\perp^2} \bigg )^2~.
\label{eq:B2pQQb} 
\end{eqnarray}

If the azimuthal angle of the final lepton $\phi_{\ell}$ is not measured,
only one of the azimuthal modulations in Eq.~(\ref{eq:cscomplete}) can be 
defined, and the cross section will be given by \cite{Boer:2011kf}\footnote{Note that the flux factor of the cross section in Eq.~(2) 
of Ref.~\cite{Boer:2011kf} has been corrected. The results on azimuthal asymmetries remain the same.}
\begin{eqnarray}
\frac{\d\sigma}
{\d y_1\,\d y_2\,\d y\,\d\xB\,\d^2\bm{q}_{\sT} \d^2\bm{K}_{\perp}} & = &
\frac{\alpha^2\alpha_s}{ \pi s M_\perp^2}\, \frac{1}{ \xB y^2}\, 
\bigg[ A + 
 B \, \bm q_\sT^2 \, \cos 2 (\phi_\perp-\phi_\sT) \bigg] \,
\delta(1 - z_1 - z_2)\,,
\label{eq:cso}
\end{eqnarray}
where we have defined $A\equiv A_0$ and $B\equiv B_0$. Further integration over
 $y_2$ leads to
\begin{eqnarray}
\frac{\d\sigma}
{\d y_1\,\d y\,\d\xB\,\d^2\bm{q}_{\sT} \d^2\bm{K}_{\perp}} & = &
\frac{\alpha^2\alpha_s}{\pi s M_\perp^2 }\, \frac{1}{\xB y^2 z(1-z)} \, 
\bigg[ A + B \, \bm q_\sT^2 \,\cos 2 (\phi_\perp-\phi_\sT) \bigg]~.
\label{eq:cso2}
\end{eqnarray}

The proposed observables involve heavy quarks in the final state, therefore 
they could be measured at high energy colliders such as the Large Hadron 
electron Collider (LHeC) proposed at CERN or at a future Electron-Ion Collider 
(EIC). The measurement or reconstruction of the
transverse momenta of the heavy quarks is essential. The individual heavy 
quark transverse momenta
$K_{i\perp}$ need to be reconstructed with an accuracy better than the
magnitude of the sum of the transverse momenta $K_{1\perp}+K_{2\perp}=q_\sT$,
which means one has to satisfy 
$\delta K_\perp \ll \vert q_{\scriptscriptstyle T}\vert \ll 
\vert K_\perp\vert$, requiring a sufficiently large $\vert K_\perp\vert$.

\begin{figure*}[t]
\begin{center}
 \includegraphics[angle=0,width=0.48\textwidth]{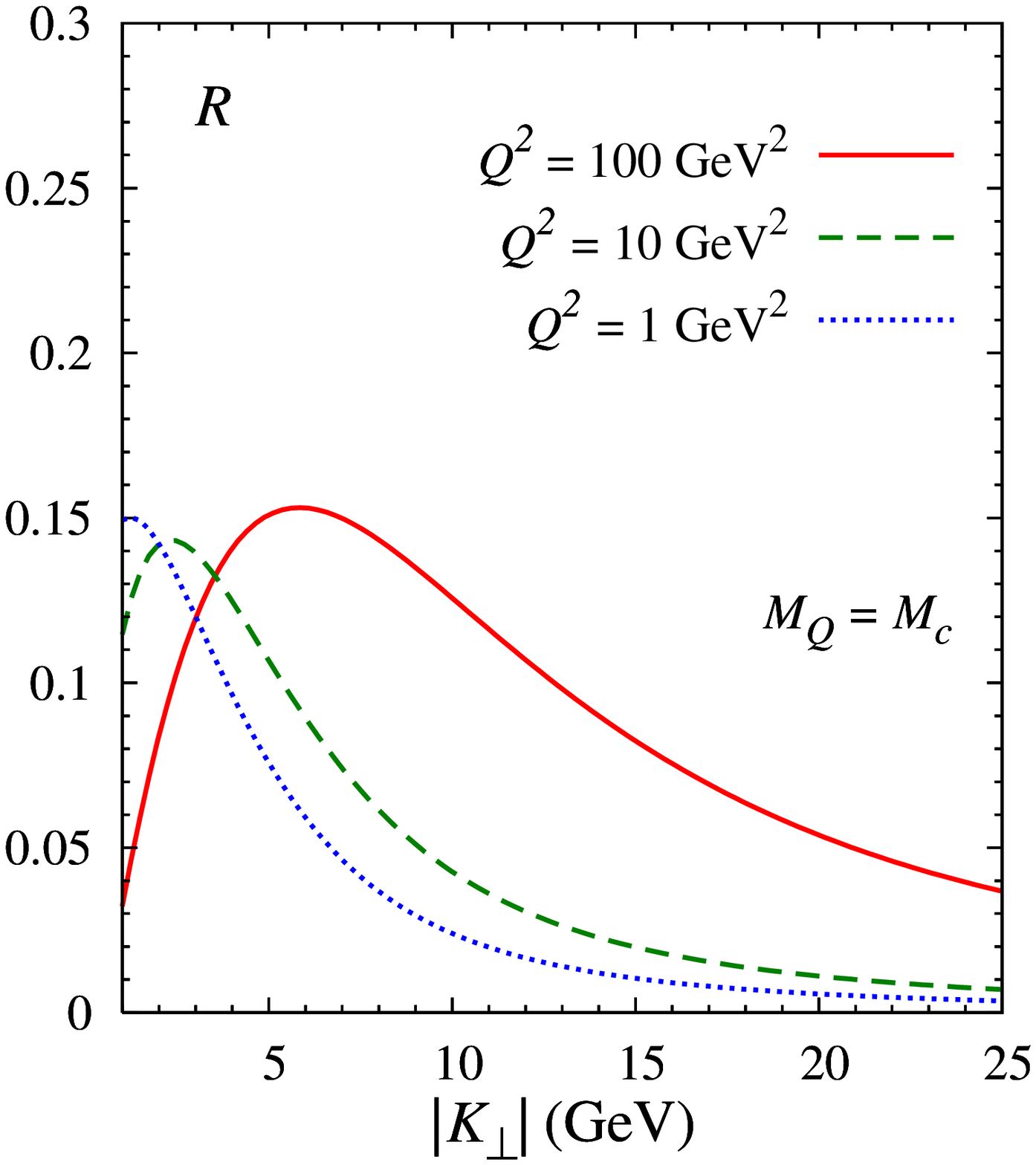}
 \includegraphics[angle=0,width=0.48\textwidth]{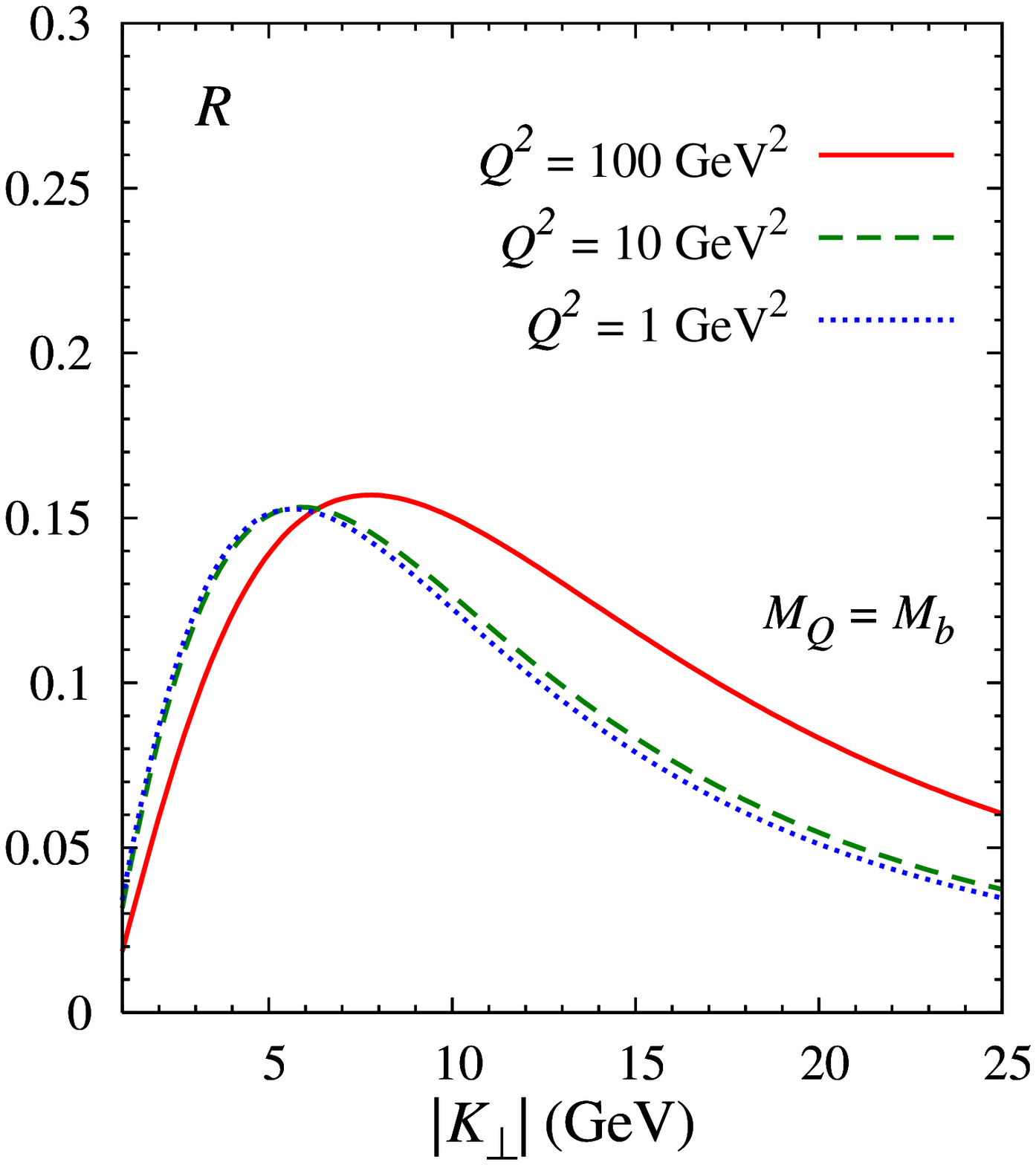}
 \caption{Upper bounds $R$ on $\vert \langle\cos 2 (\phi_\perp-\phi_\sT)\rangle \vert $  plotted as a 
function of $\vert \boldsymbol K_\perp\vert$ ($>$ 1 GeV)
at different values of 
$Q^2$ for charm (left panel) and bottom (right panel)
production in the process $e h\to e^\prime Q \bar{Q} X$, calculated at $z=0.5$, 
$y=0.01$.}
\label{fig:asy} 
\end{center}
\end{figure*}
\begin{figure*}[t]
\begin{center}
 \includegraphics[angle=0,width=0.48\textwidth]{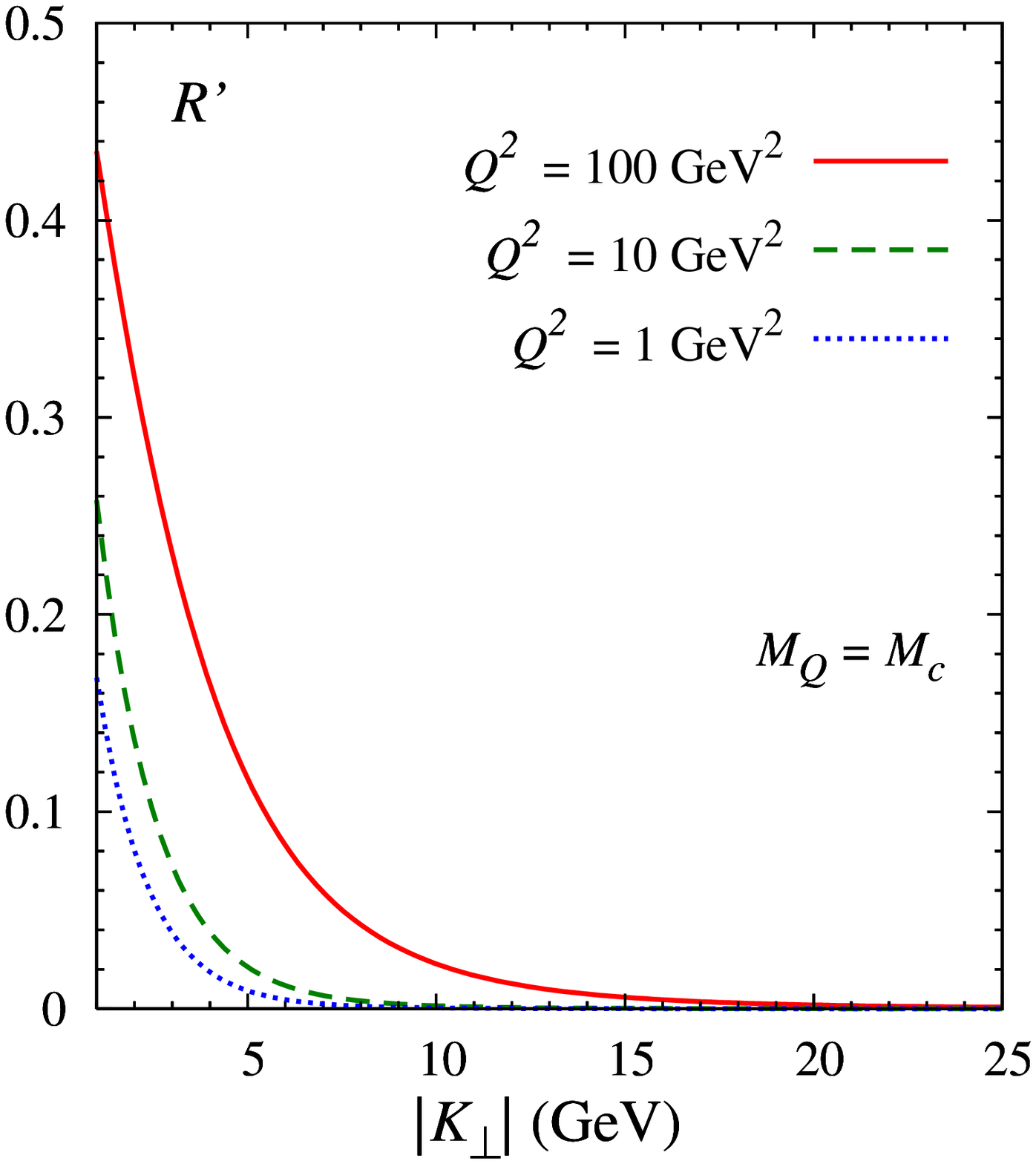}
 \includegraphics[angle=0,width=0.48\textwidth]{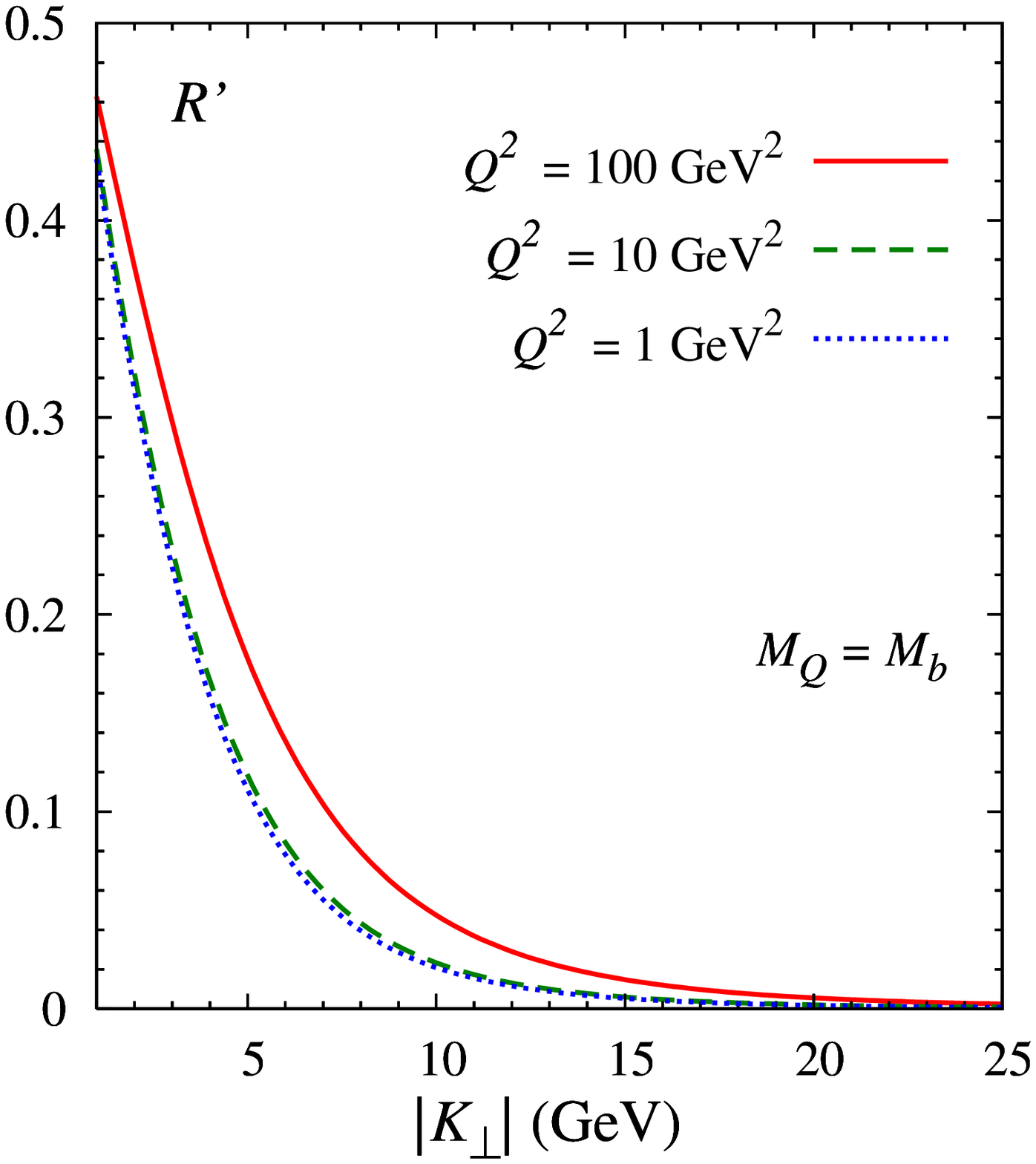}
 \caption{Same as in Fig.\ \ref{fig:asy}, but for the upper bounds $R'$  on
$\vert \langle \cos 2 \phi_{\scriptscriptstyle T} \rangle \vert $.}
\label{fig:asyp} 
\end{center}
\end{figure*}

One observes from Eqs.~(\ref{eq:BQQbphi})-(\ref{eq:B0LQQb}) that 
the magnitude $B_0$ of the
 $\cos 2 (\phi_\perp -\phi_\sT)$ modulation in Eq.~(\ref{eq:cscomplete}) 
is determined by $h_1^{\perp\, g}$
and that if $Q^2$ and/or $M_Q^2$ are of the same order as
$K_\perp^2$, the coefficient $B_0$ is not power suppressed. Using the 
positivity bound~\cite{Mulders:2000sh}
\begin{equation}
\frac{\bm p_\sT^2}{2 M^2}\,|h_1^{\perp\, g}(x,\bm p_\sT^2 )| \leq f_1^g(x,\bm p_\sT^2 )\, ,
\label{bound}
\end{equation}
we arrive at the maximum value $R$ on $| \langle \cos 2 (\phi_\perp-\phi_\sT) \rangle |$: 
\begin{equation}
\vert \langle \cos 2 (\phi_\perp-\phi_\sT) \rangle \vert = \left| \frac{\int 
\d \phi_\perp\d \phi_\sT
\, \cos 2 (\phi_\perp-\phi_\sT) \, \d\sigma}{\int \d \phi_\perp \d \phi_\sT
\, \d\sigma}\right| = \frac{ \bm{q}_\sT^2\, | B_0|}{
2 \, A_0} = \frac{\bm q_\sT^2}{2 M^2}\,\frac{|h_1^{\perp\, g}(x,\bm p_\sT^2 )|}{ f_1^{g}(x,\bm p_\sT^2 )} \, \frac{\vert {\cal B}_0 \vert}{{\cal A}_0} \le
\frac{\vert {\cal B}_0 \vert}{{\cal A}_0} \equiv R \, .
\label{eq:R}
\end{equation}
The upper bound $R$ is depicted in Fig.\ 
\ref{fig:asy} as a function of $\vert \boldsymbol K_\perp\vert$ ($>$ 1 GeV) 
at different values of 
$Q^2$ for charm (left panel) and bottom (right panel)
production, where we have selected $y= 0.01$, $z=0.5$, and taken $M_c^2=$ 2 
GeV$^2$, $M_b^2=$ 25 GeV$^2$. 
Asymmetries of this size, together with the relative simplicity of the 
suggested measurement (polarized beams are not
required), likely will allow an extraction of $h_1^{\perp\, g}$ at EIC 
(or LHeC). The bound $R'$ on
$|\langle \cos 2 \phi_\sT\rangle |$ is similarly defined:
\begin{equation}
\vert \langle \cos 2 \phi_\sT \rangle \vert = \left| \frac{\int 
\d \phi_\perp\d \phi_\sT
\, \cos 2 \phi_\sT \, \d\sigma}{\int \d \phi_\perp \d \phi_\sT
\, \d\sigma}\right| = \frac{ \bm{q}_\sT^2\, | B_2|}{
2 \, A_0} = \frac{\bm q_\sT^2}{2 M^2}\,\frac{|h_1^{\perp\, g}(x,\bm p_\sT^2 )|}{ f_1^{g}(x,\bm p_\sT^2 )} \, \frac{\vert {\cal B}_2 \vert}{{\cal A}_0} \le
\frac{\vert {\cal B}_2 \vert}{{\cal A}_0} \equiv R^\prime \, ,
\end{equation}
and is shown in Fig.\ \ref{fig:asyp} in the same kinematic
region as in Fig.\ \ref{fig:asy}. One can see 
that $R'$ can be larger than $R$, but only at smaller $\vert \boldsymbol K_\perp\vert$. $R'$ falls off more rapidly at larger 
values of $\vert \boldsymbol K_\perp\vert$ than $R$. 

Finally, we point out that final state heavy quarks can also arise from diagrams where intrinsic charm or bottom quark pairs couple to two or more valence quarks \cite{Brodsky:1980pb,Brodsky:1981se,Pumplin:2005yf,Pumplin:2007wg,Chang:2011vx,Chang:2011du}, thus contributing primarily in the valence region ($x>0.1$).  Therefore, the expressions for heavy quark pairs created in the photon-gluon fusion process, as presented in this paper, should be applicable for smaller $x$ values, which means $s \gg M_\perp^2, Q^2$.
Moreover, the intrinsic probabilities scale as $1/M^2_Q$, unlike the logarithmic contributions from gluon splitting.  Strong polarization correlations of the intrinsic heavy quarks are possible because of their multiple couplings to the projectile hadron. This is clearly worth further investigation.

\subsection{Dijet production} 

The cross section for the process 
\begin{equation}
e(\ell) + h(P) \to e(\ell^{\prime}) + {\rm jet}(K_1) + {\rm jet}(K_2)+ X
\end{equation}
can be calculated in the same way as previously described for heavy quark
 production. This means that Eqs.\ 
(\ref{eq:qexpansion})-(\ref{eq:cscomplete}) and Eqs.~(\ref{eq:cso})-(\ref{eq:cso2}) still hold when $M_Q=0$. One can
then also replace the rapidities of the outgoing particles, $y_i$, with 
the pseudo-rapidities 
$\eta_i {=}\,{-}\ln\big [\tan(\frac{1}{2}\theta_i)\big]$, 
$\theta_i$ being the polar angles of the final partons in the 
virtual photon-hadron center of mass frame. 
The explicit expressions for $A_i$, $B_i$, $B_{1,2}^\prime$ appearing in Eq.~(\ref{eq:cscomplete}) are given below. Note that $A_i$ now receive contributions from two subprocesses, namely
$e q\to e^{\prime} q g$ and $e g\to e^{\prime} q \bar{q}$. Therefore
the upper bounds of the asymmetries will be smaller than the ones for
heavy quark pair production presented in the previous section. 
More explicitly, one can write 
\begin{equation}
A_i^{e h\to e \,{\rm jet}\, {\rm jet } X}
 = \sum_{q, \bar q} e_q^2 \,C_F\,{\cal A}_i^{e q\to e q g} \, f_1^q (x,\bm{q}_{\sT}^2) + \sum_q e_q^2 \,T_R\,{\cal A}_i^{e g\to e q \bar q} \, f_1^g (x,\bm{q}_{\sT}^2)\, , \qquad i =1,2,3\,,
\label{eq:A}
\end{equation}
where $C_F = (N_c^2-1)/2 N_c$, with $N_c$ being the number of colors, and, similarly to Eq.~(\ref{eq:Agstar}),
\begin{eqnarray}
{\cal A}_0^{e q\to e q g} & = & [1+(1-y)^2]\,{\cal A}_{U+L}^{\gamma^* q\to q g} \,
-y^2 \,{\cal A}_{L}^{\gamma^* q\to q g}\,, \nonumber \\
{\cal A}_1^{e q\to q g} & = & (2-y) \sqrt{1-y}\,{\cal A}_{I}^{\gamma^* q\to q g}\,,\nonumber \\
 {\cal A}_2^{e q\to e q g} & = & 2 (1-y)\, {\cal A}_{T}^{\gamma^* q\to q g}\,,\label{eq:Aqstar} 
\end{eqnarray}
see also Eq.~(\ref{eq:Ahel}). Neglecting terms suppressed by powers of 
$|\bm{q}_\sT| /|\bm K_\perp|$, in agreement with the results in Ref.~\cite{Mirkes:1997uv}, we obtain
\begin{eqnarray}
{\cal A}_{U+L}^{\gamma^* q\to q g} & = & \frac{1-z}{ D_0^2}\, 
\left\{ 1+z^2 + \left [ 2 z(1-z) + 4z^2(1-z)^2\right ] \,\frac{Q^2}{\bm K_\perp^2} + \left [z^2 (1-z)^2\right] \left [ 1+(1-z)^2\right]\, \frac{Q^4}{\bm K_\perp^4}\right \}\,, \\
{\cal A}_{L}^{\gamma^* q\to q g} & = & 4 \,\frac{z^2(1-z)^3}{D_0^2}\,\frac{Q^2}{\bm K_\perp^2} \,,\\
{\cal A}_I^{\gamma^* q\to q g} & = & 
- 4 \,\frac{z^2(1-z)^2}{D_0^2}\,\left [ 1+ (1-z)^2 \,\frac{Q^2}{\bm K_\perp^2} \right ] \,\frac{Q}{\vert \bm K_\perp \vert}\, ,\\
{\cal A}_{T}^{\gamma^* q\to q g} & = & 
\,2 
\frac{z^2(1-z)^3}{D_0^2}\, \frac{Q^2}{\bm K_\perp^2}\,,
\label{eq:A2ej}
\end{eqnarray}
with
\begin{equation}
D_0 \equiv D_0 \left (z, \frac{Q^2}{\bm K_\perp^2} \right ) = 1 + z (1-z) \frac{Q^2}{\bm K_\perp^2}~.
\label{eq:Den0}
\end{equation}
Furthermore, taking $M_Q=0$ and $M_\perp = |\bm K_\perp|$ in Eqs.~(\ref{eq:AQQb2})-(\ref{eq:Den}), for the subprocess $\gamma^* g \rightarrow q \bar{q}$ we get
\begin{eqnarray}
{\cal A}_{U+L}^{\gamma^* g\to q \bar q} & = &\frac{1}{D_0^3}\, 
\left \{ 1-2 z (1-z) + z^2 (1-z)^2 \left 
[ 8 \frac{Q^2}{\bm K_\perp^2} + [1- 2z (1-z)]\, \frac{Q^4}{\bm K_\perp^4} \right ] \right \} \,, \\
{\cal A}_{L}^{\gamma^* g\to q \bar q} & = & 8\, \frac{z^2(1-z)^2}{D_0^3}\,
\frac{Q^2}{\bm K_\perp^2} \, ,\\
{\cal A}_{I}^{ \gamma^* g \to q \bar q} & = & 4\,
\frac{z(1-z)(1-2 z)}{D_0^3}\,
\bigg [1- z (1-z) \frac{Q^2}{\bm K_\perp^2} \bigg ]\, 
\frac{Q}{|\bm K_\perp|}\,, \\
{\cal A}_{T}^{\gamma^* g\to e q \bar q} & = & 4 \,
\frac{z^2(1-z)^2}{D_0^3}\, \frac{Q^2}{\bm K_\perp^2}~.
\end{eqnarray}

In analogy to Eq.~(\ref{eq:BQQbphi}), we have for the terms that depend
on the gluon distribution function $h_1^{\perp\, g}$: 
\begin{eqnarray}
B_i^{e h\to e \,{\rm jet}\, {\rm jet } X} & = & 
\frac{1}{M^2}\,\sum_q e^2_q \,{\cal B}_i^{e g\to e q \bar q} \, h_1^{\perp \,g} (x, \bm{q}_{\sT}^2)\, , \nonumber 
\label{eq:B} \\
B_{1, 2}^{\prime \,e h\to e {\rm jet }\, {\rm jet} X} & = & 
\frac{1}{M^2}\,\sum_q e^2_q\,{\cal B}_{1, 2}^{\prime \,e g\to e {q} \bar q }\, h_1^{\perp \,g} (x, \bm{q}_{\sT}^2)\, ,
\end{eqnarray}
with
\begin{eqnarray}
{\cal B}_0^{e g\to e q \bar q} & = & [1+(1-y)^2]\,{\cal B}_{U+L}^{\gamma^* g\to q \bar q} 
-y^2 \,{\cal B}_{L}^{\gamma^* g\to q \bar q}\,, \nonumber \\
{\cal B}_1^{e g\to e q \bar q} & = & (2-y) \sqrt{1-y}\,{\cal B}_{I}^{\gamma^* g\to q \bar q}\,, \hspace{2.2cm} {\cal B}_1^{\prime \,e g\to e q \bar q} \, = \, (2-y) \sqrt{1-y}\,{\cal B}_{I}^{\prime\,\gamma^* g\to q \bar q}\,, \nonumber \\
 {\cal B}_2^{e g\to e q \bar q} & = & 2 (1-y)\, {\cal B}_{T}^{\gamma^* g\to q \bar q}\,,\hspace{3.18cm} {\cal B}_2^{\prime\, e g\to e q \bar q} \, = \, 2 (1-y)\, {\cal B}_{T}^{\prime\, \gamma^* g\to q \bar q}~. 
\end{eqnarray}
By taking $M_Q=0$ and $M_\perp = |\bm K_\perp|$ in Eqs.~(\ref{eq:B0QQb})-(\ref{eq:B2pQQb}), we obtain
\begin{eqnarray}
{\cal B}_{U+L}^{\gamma^* g\to q \bar q} & = &-\frac{z(1-z)[1-6z(1-z)]}{D_0^3}\, \frac{Q^2}{\bm K_\perp^2} \,, \\
{\cal B}_{L}^{\gamma^* g\to q \bar q} & = & 4\, \frac{z^2(1-z)^2}{D_0^3}\,
\frac{Q^2}{\bm K_\perp^2} \, ,\\
{\cal B}_{I}^{\gamma^* g\to q\bar q} & = & - 2 \,
\frac{z^2(1-z)^2(1-2 z)}{D_0^3}\,\frac{Q^3}{\vert \bm K_\perp\vert^3}
\, ,\\
{\cal B}_{I}^{\prime\,\gamma^* g\to q\bar q} & = & 2 \,
\frac{z(1-z)(1-2 z)}{D_0^3}\, \frac{Q}{|\bm K_\perp|}\, ,\\
{\cal B}_{T}^{\gamma^* g\to q\bar q} & = & - 
\frac{z^3(1-z)^3}{D_0^3}\,\frac{Q^4}{\bm K_\perp^4} \, ,\\
{\cal B}_{T}^{\prime\, \gamma^* g\to q\bar q} & = & -
\frac{z(1-z)}{D_0^3}~.
\end{eqnarray}

\subsection{Dilepton production}

Azimuthal modulations analogous to the ones calculated above arise in QED as well, in the `tridents' processes $\ell e(p) \to \ell \mu^+ \mu^- e^\prime (p^\prime\,{\rm or}\, X)$ or
$\mu^- Z \to \mu^- \ell \bar{\ell} Z$ \cite{Bjorken:1966kh,Brodsky:1966vh,Tannenbaum:1968zz,Gluck:2002fi,Gluck:2002cm}. Such asymmetries could be described by the distribution of linearly polarized photons inside a lepton, proton, or atom. The 
transverse momentum dependent unpolarized and linearly polarized photon
distributions in a hadron, denoted by $f_1^{\gamma}(x, \bm p_\sT^2)$ and 
$h_1^{\perp\,\gamma}(x, \bm p_\sT^2)$ respectively, can be defined in the 
same way as their gluonic counterparts, see Eq.~(\ref{eq:gcorr}). Therefore, 
the cross section for the electroproduction of two muons, 
\begin{equation}
e (\ell){+}h(P)\to e(\ell^\prime) {+} \mu^{-}(K_1) {+} \mu^{+}(K_2){+}X \, ,
\end{equation}
proceeds, at LO in QED, via the subprocess
\begin{equation}
\gamma^*(q) + \gamma(p) \to \mu^{-}(K_1) {+} \mu^{+}(K_2)\,,
\end{equation}
where the second (real) photon is emitted by the hadron. If the $\mu^- \mu^+$ pair in the final state is almost back-to-back in the
plane perpendicular to the direction of the exchanged (virtual) photon and hadron, the corresponding cross section is the same as the one in
 Eq.~(\ref{eq:cscomplete}) derived for $Q\bar Q$ production, with $\alpha_s$ replaced by $\alpha$ and $M_Q$ by $M_\mu$. The coefficients of the various azimuthal 
modulations are those given in Eqs.~(\ref{eq:AQQb})-(\ref{eq:B2pQQb}) with the replacements $e_Q^2 \to 1$, $T_R\to 1$, $f_1^g\to f_1^{\gamma}$, $h_1^{\perp\,g} \to h_1^{\perp\,\gamma}$.

\begin{figure*}[t]
\begin{center}
 \includegraphics[angle=0,width=0.48\textwidth]{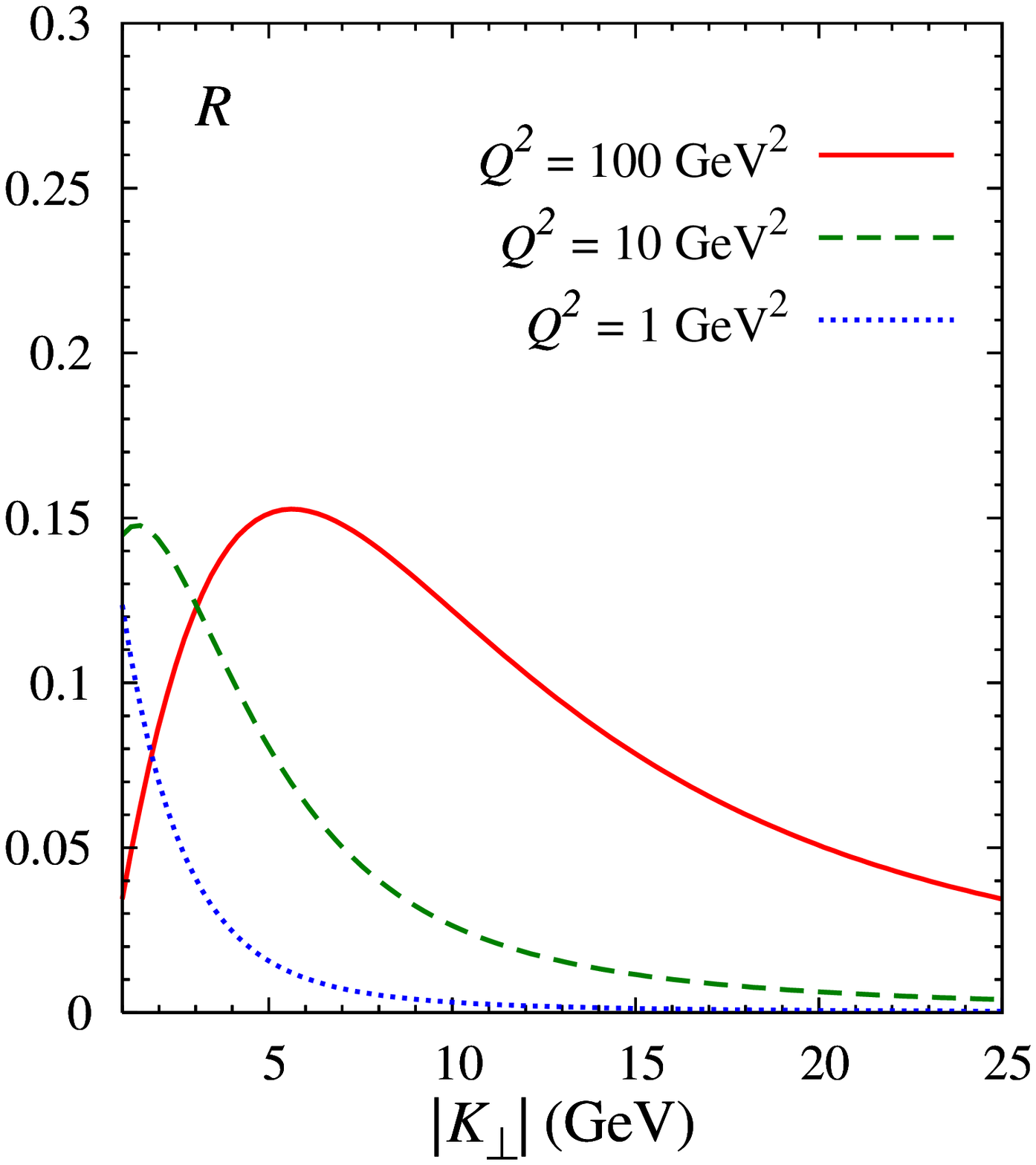}
 \includegraphics[angle=0,width=0.48\textwidth]{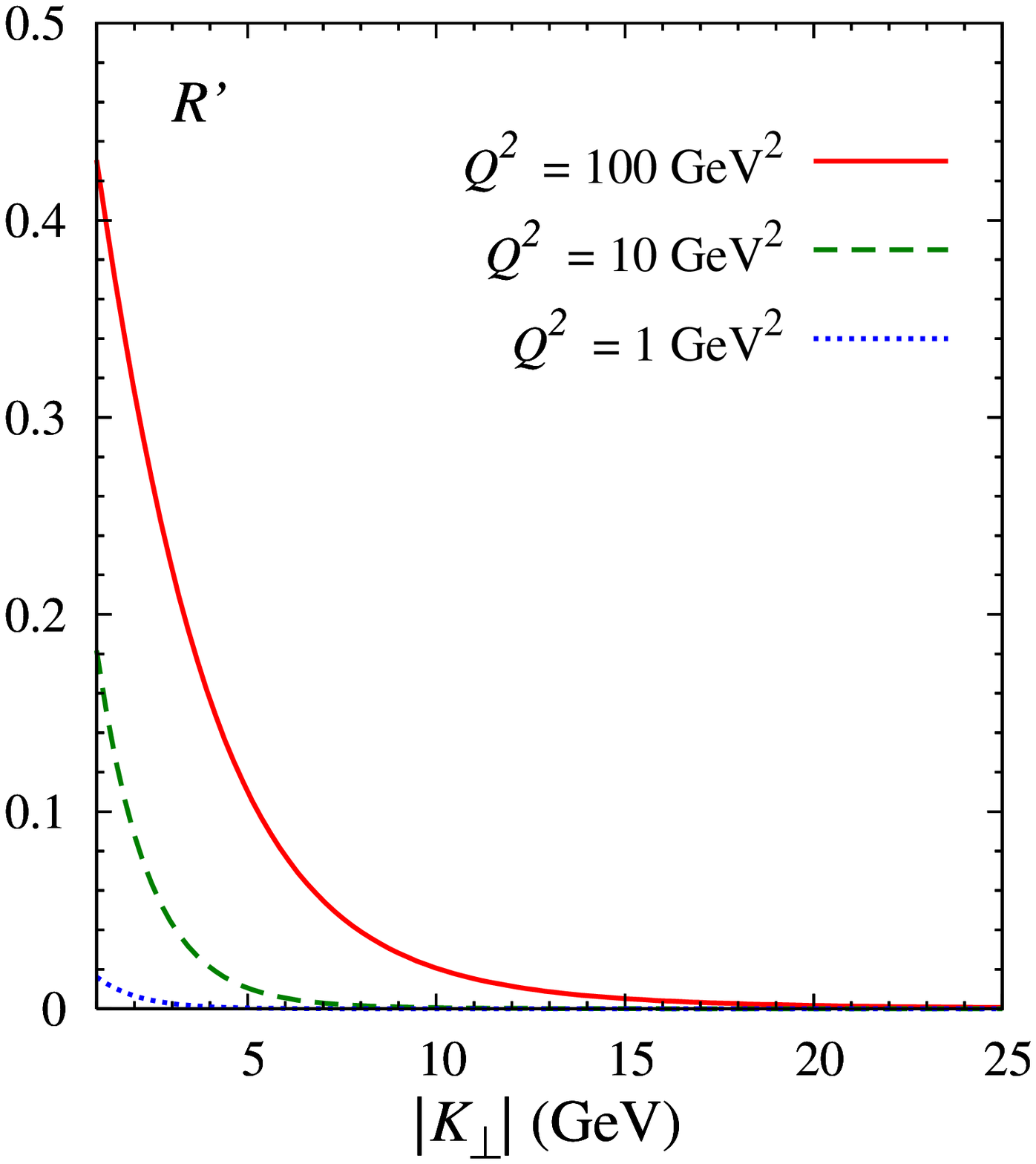}
 \caption{Upper bounds $R$ (left panel) and $R^\prime$ (right panel) on $\vert \langle \cos 2 (\phi_\perp-\phi_\sT) \rangle\vert$ and $\vert \langle 
\cos 2 \phi_\sT \rangle \vert $, respectively, as a
 function of $\vert \boldsymbol K_\perp\vert$  ($>$ 1 GeV)
 at different values of 
 $Q^2$, for the process $e h\to e^\prime \mu^- \mu^+ X$, 
calculated at $z=0.5$, 
$y=0.01$.
\label{fig:asy_gamma} }
\end{center}
\end{figure*}

The bounds $R$ and $R^\prime$ for the process $e h\to e^\prime \mu^- \mu^+ X$ can be obtained using the positivity constraint for linearly polarized photon distributions, analogous to the one in Eq.~(\ref{bound}) for gluons, and 
they are shown in Fig.\ \ref{fig:asy_gamma}. Especially as $Q^2$ increases, they become very similar to $R$ and $R^\prime$ for the process $e h\to e^\prime Q \bar{Q} X$.

\section{Hadron-hadron collisions}
\subsection{Heavy quark production}

The cross section for the process
\begin{equation}
h_1(P_1){+}h_2(P_2)\, {\rightarrow}\,Q(K_1){+} \bar{Q}(K_2){+}X\, , 
\end{equation}
in a way similar to the hadroproduction of two jets discussed in Ref.~\cite{Boer:2009nc} (to which we refer for the details of the calculation),
can be written in the following form
\begin{equation}
\frac{\d\sigma}
{\d y_1 \d y_2 \d^2 \bm{K}_{1\perp} \d^2 \bm{K}_{2\perp}} =
\frac{\alpha_s^2}{s {M}_\perp^2}
\bigg[ A(\bm{q}_\sT^2) + B(\bm{q}_\sT^2) \bm q_\sT^2 \cos 2 (\phi_\perp - \phi_\sT) + 
C(\bm{q}_\sT^2) \bm{q}_\sT^4 \cos 4 (\phi_\perp-\phi_\sT)\bigg ]\, ,
\label{eq:csoQQb}
\end{equation}
where $y_i$ are the rapidities of 
the outgoing particles, $\bm{q}_\sT 
\equiv \bm{K}_{1 \perp} + \bm{K}_{2 \perp}$, 
 $\bm{K}_\perp \equiv (\bm{K}_{1\perp}
- \bm{K}_{2\perp})/2$ and $M_\perp = 
\sqrt{M_Q^2 + \bm K_{\perp}^2}$, $M_Q$ being the heavy quark mass. 
The momentum $\bm{q}_\sT$ is in principle experimentally accessible
and is related to the intrinsic transverse momenta
of the incoming partons,
$\bm{q}_\sT = \bm{p}_{1 T} + \bm{p}_{2 T}$. The azimuthal angles of 
$\bm{K}_\perp$ and $\phi_\sT$ are denoted by $\phi_\sT$ and $\phi_\perp$,
respectively. 
Besides $\bm q_\sT^2$, the terms $A$, $B$ and $C$ depend 
on other kinematic variables not explicitly shown, 
such as $z$, which is given in Eq.~(\ref{Yexpression}) with $Q^2 = 0$
 and with the Mandelstam variables defined by the momenta of the 
incoming ($p_1$, $p_2$) and outgoing ($K_1$, $K_2$) partons as follows,
\begin{equation}
\hat s = (p_1 + p_2)^2, \qquad \hat t = (p_1-K_1)^2 ,\qquad 
\hat u = (p_1-K_2)^2~.
\end{equation}
Furthermore, they depend on $M_Q^2/M_\perp^2$ and on the 
light-cone 
momentum fractions $x_1$, $x_2$, related to the rapidities, 
the mass and the transverse momenta of the heavy quark and antiquark by the
 relations 
\begin{equation} 
x_1 =\frac{1}{\sqrt s}\bigg(\,M_{1 \perp}\,e^{y_1}\,
{+} M_{2 \perp}\,e^{y_2}\,\bigg ), \,\quad
x_2{=}\frac{1}{\sqrt s}
\bigg (\, M_{1 \perp}e^{-y_1}\,
{+} M_{2 \perp}\,e^{-y_2}\, \bigg )\, , 
\end{equation}
with, as before, 
$M^2_{i \perp} = \bm K_{ i \perp}^2 + M_Q^2 \approx M_\perp^2$.

The terms $A$, $B$, and $C$ have been calculated at LO in perturbative QCD, 
adopting the approximation 
$|\bm{q}_\sT| \ll |\bm{K}_{1 \perp}| \approx
|\bm{K}_{2 \perp}| \approx |\bm{K}_\perp|$ 
which is applicable when the heavy quark and antiquark pair is produced 
 almost back-to-back in the transverse plane. Their explicit expressions, 
which contain convolutions of different TMDs, are given in the following. 
As discussed in Ref.~\cite{Boer:2010zf}, the coefficients $B$ and $C$ 
in Eq.~(\ref{eq:csoQQb}) could be separated by $\bm{q}_\sT^2$-weighted 
integration over $\bm{q}_\sT$. We point out that in the limiting 
situation when $|\bm{K}_{1 \perp}| = |\bm{K}_{2 \perp}|$, one has exactly 
$\cos 2 (\phi_\perp-\phi_\sT)= -1$ and 
$\cos 4 (\phi_\perp-\phi_\sT)=1$, since $K_\perp$ and $q_\sT$ are orthogonal. 
In this case the remaining angular dependence (on the imbalance angle $\delta \phi = \phi_Q -\phi_{\bar{Q}}-\pi$) 
enters through $\bm{q}_\sT^2$ only \cite{Boer:2010zf}.
 
The angular independent part $A$ of the cross section in 
Eq.~(\ref{eq:csoQQb}) is given by the sum 
of the contributions ${\cal A}^{q \bar q \to Q \bar{Q}}$ and 
${\cal A}^{g \bar g \to Q \bar{Q}}$, coming respectively from the 
partonic subprocesses $q \bar q\to Q \bar{Q}$ and $g g \to Q \bar{Q}$, 
which underlie the process $h_1 h_2\to Q \bar{Q} X$:
\begin{equation}
A = {\cal A}^{q \bar q \to Q \bar Q} + {\cal A}^{g g \to Q \bar Q}\,,
\label{eq:Ajet}
\end{equation}
with
\begin{eqnarray}
{\cal A}^{q \bar{q} \to Q \bar Q} &= & \frac{N_c^2-1}{2 N_c^2}\,z (1- z) \,
\left [ {z}^2 + (1- z)^2 + 2 z (1- z) \, \frac{M_Q^2}{M_\perp^2} \right ] \bigg [{\cal{F}}^{q \bar{q}}(x_1, x_2, \bm{q}_{\sT}^2)+ {\cal{F}}^{\bar q q}(x_1, x_2, \bm{q}_{\sT}^2) \bigg ]\, , \label{eq:Aqqb}\\
{\cal A}^{g g \to Q \bar{Q} } & =& {\cal A}_{1} \left (z, \frac{M_Q^2}{M_\perp^2} \right ) {\cal{F}}^{g g} (x_1, x_2,\bm q_\sT^2) + \frac{M_Q^4}{M_\perp^4}\,
{\cal A}_{2} (z)\, \bm q_\sT^4\,{\cal N}^{g g}(x_1, x_2,\bm q_\sT^2)\, ,
\label{eq:Agg2}
\end{eqnarray} 
where 
\begin{eqnarray}
{\cal A}_{1}& =& \frac{N_c}{N_c^2-1}\,\frac{1}{2} \bigg ({z} ^2 + (1-z)^2 - \frac{1}{N_c^2}\bigg ) \left [ {z}^2 + (1-z)^2 + 4 z (1- z)\left (1-\frac{M_Q^2}{M_\perp^2} \right )\frac{M_Q^2}{M_\perp^2}\right ] \, ,\label{eq:A1} \\
{\cal A}_2 & = & - \frac{N_c}{N_c^2-1}\,\frac{z (1-z)}{4} \, 
\left [z^2 + (1-z^2)-\frac{1}{N_c^2} \right ]~. 
\label{eq:A2}
\end{eqnarray}
We have adopted the following convolutions of TMDs,
\begin{eqnarray}
{\cal{F}}^{a b}(x_1, x_2, \bm{q}_{\sT}^2) & \equiv & 
\int d^2\bm{p}_{1 \sT}\,d^2
\bm{p}_{2 T} 
\,\delta^2 (\bm{p}_{1 \sT} +\bm{p}_{2 T} -\bm{q}_{\sT})
 f_1^{a}(x_1, \bm p^2_{1 \sT}) f_1^{{b}} 
(x_2, \bm p^2_{2 T})\, , 
\label{eq:Fqq}
\end{eqnarray}
where a sum over all (anti)quark flavors is understood, and 
\begin{eqnarray}
\bm q_\sT^4 \,{\cal{N}}^{g g}(x_1, x_2, \bm{q}_{\sT}^2) & \equiv & 
\frac{1}{M_1^2 M_2^2}\int d^2\bm{p}_{1 \sT}\,d^2
\bm{p}_{2 \sT} 
\,\delta^2 (\bm{p}_{1 \sT} +\bm{p}_{2 \sT} -\bm{q}_{\sT})
\left [ 2 (\bm p_{1 \sT}\cdot \bm p_{2 \sT})^2 - 
\bm p_{1 \sT}^2 \bm p_{2 \sT}^2 \right ] \nonumber \\
&&\qquad \times h_{1}^{\perp g}(x_1, \bm p^2_{1 \sT}) h_1^{{\perp g }} 
(x_2, \bm p^2_{2 \sT})~.
\label{eq:Ngg}
\end{eqnarray}
The results in 
Eq.~(\ref{eq:Aqqb}) and in Eq.~(\ref{eq:Agg2}), 
integrated over $\bm q_\sT$, 
recover the ones calculated in the framework of collinear LO 
pQCD, which can be found, for example, in Refs.~\cite{Kniehl:2004fy,Gluck:1977zm,Anselmino:2004nk} and in Refs.~\cite{Kniehl:2004fy,Anselmino:2004nk}, respectively. Moreover, taking the limit $M_Q\to 0$, agreement is found between 
 Eqs.~(\ref{eq:Aqqb})-(\ref{eq:A2}) and the explicit expressions derived 
for massless partons published in Ref.~\cite{Boer:2009nc} [Eqs.~(23), (28)],
namely
\begin{eqnarray}
{\cal A}^{q \bar{q} \to q' \bar q'} &= & \frac{N_c^2-1}{2 N_c^2}\, z (1-z) \,
\left [z^2 + (1-z)^2 \right ]\, \bigg [{\cal{F}}^{q \bar{q}}(x_1, x_2, \bm{q}_{\sT}^2)+ {\cal{F}}^{\bar q q}(x_1, x_2, \bm{q}_{\sT}^2) \bigg ] \, , 
\end{eqnarray}
and
\begin{eqnarray}
{\cal A}^{g g \to q \bar{q} } & =& \frac{N_c}{N_c^2-1}\, \bigg [z^2 + (1-z)^2 - \frac{1}{N_c^2}\bigg ] \,
\frac{z^2 + (1-z)^2}{2}\, {\cal{F}}^{g g} (x_1, x_2,\bm q_\sT^2)~.
\end{eqnarray}

In analogy to Eq.~(\ref{eq:Ajet}), we write
\begin{equation}
B = {\cal B}^{q \bar q \to Q \bar{Q}} 
+ 
\frac{M_Q^2}{M_\perp^2} \, {\cal B}^{g g \to Q \bar{Q}} \, , 
\label{eqBjet}
\end{equation}
where
\begin{eqnarray}
{\cal B}^{q \bar q \to Q\bar Q } & = & \frac{N_c^2-1}{N_c^2}\,{z}^2 (1-z)^2 \left (1 -\frac{M_Q^2}{M_\perp^2} \right )\bigg [ {\cal{H}}^{q \bar{q}}(x_1, x_2,\bm{q}_{\sT}^2) +
 {\cal{H}}^{\bar q q}(x_1, x_2,\bm{q}_{\sT}^2) \bigg ]\, ,\\
{\cal B}^{g g \to Q\bar Q } & = & \frac{N_c}{N_c^2-1}\,{\cal B}_1 \left (z, 
\frac{M_Q^2}{M_\perp^2} \right ){\cal{H}}^{g g }(x_1, x_2,\bm{q}_{\sT}^2)\,,
\end{eqnarray}
with
\begin{eqnarray}
{\cal B}_1 & = & {z} (1-z) 
\left [{z}^2 +(1-z)^2 -\frac{1}{N_c^2}\right ] 
\left (1 -\frac{M_Q^2}{M_\perp^2} \right )\,.
\label{eq:B1}
\end{eqnarray}

Similarly to Eqs.~(\ref{eq:Fqq}) and (\ref{eq:Ngg}), we have defined 
the following convolutions of parton distributions 
\begin{eqnarray}
\bm{q}_\sT^2\,{\cal{H}}^{q \bar q}(x_1, x_2, \bm{q}_{\sT}^2) 
& \equiv &
\frac{1}{M_1 M_2}\sum_{\rm flavors} \int d^2\bm{p}_{1 \sT}\,d^2
\bm{p}_{2 \sT} 
\,\delta^2 (\bm{p}_{1 \sT} +\bm{p}_{2 \sT} -\bm{q}_{\sT})
\nonumber \\
&& \qquad \qquad\qquad \mbox{} \times 
\left[2 (\bm{\hat{h}} \cdot \bm{p}_{1 \sT})
(\bm{\hat{h}} \cdot \bm{p}_{2 \sT}) 
- ( \bm{p}_{1 \sT}\cdot \bm{p}_{2 \sT}) \right]
h_1^{\perp q}(x_1, \bm p^2_{1 \sT}) 
h_1^{\perp \bar q} (x_2, \bm p^2_{2 \sT})\, ,
\label{eq:Hqq}
\end{eqnarray}
and
\begin{eqnarray}
\bm{q}_\sT^2\,{\cal{H}}^{g g }(x_1, x_2, \bm{q}_{\sT}^2) 
& \equiv &
\frac{1}{M_1 M_2} \int d^2\bm{p}_{1 \sT}\,d^2
\bm{p}_{2 \sT} 
\,\delta^2 (\bm{p}_{1 \sT} +\bm{p}_{2 \sT} -\bm{q}_{\sT})
\left \{ \left[2 (\bm{\hat{h}} \cdot \bm{p}_{1 \sT})^2 
- \bm{p}_{1 \sT}^2 \right] h_1^{\perp g}(x_1, \bm p^2_{1 \sT}) 
f_1^{g} (x_2, \bm p^2_{2 \sT}) \right . \nonumber \\
&&\qquad\qquad \qquad +\left . \left[2 (\bm{\hat{h}} \cdot 
\bm{p}_{2 \sT})^2 - \bm{p}_{2 \sT}^2 \right] f_1^{g} (x_1, \bm p^2_{1 \sT})
 h_1^{\perp g}(x_2, \bm p^2_{2 \sT}) \right \}\,,
\label{eq:Hgg}
\end{eqnarray}
with $\hat{\bm h}\equiv \bm q_{\sT}/ \vert \bm q_{\sT}\vert$. 
The result given in Eq.~(36) of Ref.~\cite{Boer:2009nc}, 
\begin{eqnarray}
{\cal B}^{q \bar q \to q' \bar q' } & = & \frac{N_c^2-1}{N_c^2}\, z^2 (1-z)^2 \bigg [ {\cal{H}}^{q \bar{q}}(x_1, x_2,\bm{q}_{\sT}^2) +
 {\cal{H}}^{\bar q q}(x_1, x_2,\bm{q}_{\sT}^2) \bigg ]\, ,
\end{eqnarray}
is recovered taking the massless limit of Eqs.~(\ref{eqBjet})-(\ref{eq:B1}).

Finally, the $\cos 4(\phi_\perp-\phi_\sT) $ angular distribution of the $Q\bar Q$ pair
is related exclusively to the presence of (linearly) polarized gluons inside 
unpolarized hadrons. It turns out that 
\begin{eqnarray}
 C = {\cal C}^{g g \to Q \bar Q} = {\cal C}(z) \bigg (1-\frac{M_Q^2}{M_\perp^2} \bigg )^2 \,\bigg [2 {\cal{I}}^{g g }(x_1, x_2,\bm{q}_\sT^2) - {\cal{L}}^{g g }(x_1, x_2,\bm{q}_\sT^2)\bigg ]\,,
\label{eq:Cgg} 
\end{eqnarray}
with 
\begin{equation}
{\cal C}(z) = {\cal A}_2(z)= -\frac{N_c}{N_c^2-1}\,\frac{z (1-z)}{4} \, 
\left [z^2 + (1-z^2)-\frac{1}{N_c^2} \right ]\, ,
\label{eq:Cgg2}
\end{equation}
see Eq.~(\ref{eq:A2}), where we have introduced the convolutions \cite{Boer:2009nc}
\begin{eqnarray}
\bm{q}_\sT^4\,{\cal{I}}^{g g }(x_1, x_2, \bm{q}_\sT^2 ) 
& \equiv & 
\frac{1}{M_1^2 M_2^2}\int d^2\bm{p}_{1 \sT}\,d^2
\bm{p}_{2 \sT} 
\,\delta^2 (\bm{p}_{1 \sT} +\bm{p}_{2 \sT} -\bm{q}_{\sT}) 
\nonumber \\
&& \qquad \qquad\qquad \mbox{} \times 
\bigg [2 (\hat{\bm{h}} \cdot \bm{p}_{1 \sT})
( \hat{\bm{h}} \cdot \bm{p}_{2 \sT}) 
- (\bm{p}_{1 \sT} \cdot\bm{p}_{2 \sT}) \bigg ]^2
\ h_1^{\perp g}(x_1, \bm p^2_{1 \sT}) h_1^{\perp g} 
(x_2, \bm p^2_{2 \sT}) \, ,
\label{eq:I}
\end{eqnarray}
and
\begin{equation}
\bm{q}_\sT^4\,{{\cal{L}}^{g g }(x_1, x_2,\bm{q}_\sT^2)} 
\equiv \frac{1}{M_1^2 M_2^2} 
\, \int d^2\bm{p}_{1 \sT}\,d^2
\bm{p}_{2 \sT} 
\,\delta^2 (\bm{p}_{1 \sT} +\bm{p}_{2 \sT} -\bm{q}_{\sT}) \bm p^2_{1 \sT} \bm p^2_{2 \sT}
 h_1^{\perp g}(x_1, \bm p^2_{1 \sT}) h_1^{\perp g} 
(x_2, \bm p^2_{2 \sT})~.
\label{eq:L} 
\end{equation}
In the massless limit, we recover the result in Eq.~(46) of 
Ref.~\cite{Boer:2009nc},
\begin{eqnarray}
{\cal C}^{g g \to q \bar q} = -\frac{N_c}{N_c^2-1}\,\frac{z (1-z)}{4} \, \bigg (z^2 + (1-z)^2 - \frac{1}{N_c^2}\bigg ) \, \bigg [2 {\cal{I}}^{g g }(x_1, x_2,\bm{q}_\sT^2) - {\cal{L}}^{g g }(x_1, x_2,\bm{q}_\sT^2)\bigg ]~.
\label{eq:Cqqb}
\end{eqnarray}

In arriving at the above expressions we have ignored the modifications due to initial and final state interactions. 
We address their effect in Sect.\ \ref{Colorflow}.

\subsection{Dilepton production}

The cross section for the reaction
\begin{equation}
h_1(P_1){+}h_2(P_2)\, {\rightarrow}\,\mu^-(K_1){+} \mu^+(K_2){+}X\, , 
\end{equation}
which proceeds via the two channels $q \bar q \to \mu^-\mu^+$ 
(Drell-Yan scattering) and $\gamma \gamma \to \mu^-\mu^+$ (photon fusion), can be recovered 
from the results for heavy quark pair production by taking 
the limit $N_c\to 0$ \cite{Brodsky:1997jk}. It can still be written as in Eq.~(\ref{eq:csoQQb}), with 
$\alpha_s$ replaced by $\alpha$ and 
\begin{equation}
A = {\cal A}^{q \bar q \to \mu^- \mu^+} + {\cal A}^{\gamma \gamma \to \mu^- \mu^+}\,, \quad
B = {\cal B}^{q \bar q \to \mu^- \mu^+} 
+ 
\frac{M_\mu^2}{M_\perp^2} \, {\cal B}^{\gamma \gamma \to \mu^- \mu^+} \, , \quad
C = {\cal C}^{\gamma \gamma \to \mu^- \mu^+} \, ,
\label{eq:ABCgam}
\end{equation}
with
\begin{eqnarray}
{\cal A}^{q \bar{q} \to \mu^- \mu^+} &= & 2 z (1- z) \,
\left [ {z}^2 + (1- z)^2 +2 z (1- z) \, \frac{M_\mu^2}{M_\perp^2} \right ] \bigg [{\cal{F}}^{q \bar{q}}(x_1, x_2, \bm{q}_{\sT}^2)+ {\cal{F}}^{\bar q q}(x_1, x_2, \bm{q}_{\sT}^2) \bigg ]\, , \label{eq:Amu}\\
{\cal A}^{\gamma \gamma \to \mu^- \mu^+} & =& {\cal A}_{1} \left (z, \frac{M_\mu^2}{M_\perp^2} \right ) {\cal{F}}^{\gamma \gamma} (x_1, x_2,\bm q_\sT^2) - \frac{M_\mu^4}{M_\perp^4}\,
z(1-z)\, \bm q_\sT^4\,{\cal N}^{\gamma \gamma}(x_1, x_2,\bm q_\sT^2)\, ,\\
{\cal B}^{q \bar q \to \mu^- \mu^+} & = & 4 {z}^2 (1-z)^2 \left (1 -\frac{M_\mu^2}{M_\perp^2} \right )\bigg [ {\cal{H}}^{q \bar{q}}(x_1, x_2,\bm{q}_{\sT}^2) +
 {\cal{H}}^{\bar q q}(x_1, x_2,\bm{q}_{\sT}^2) \bigg ]\, ,\\
{\cal B}^{\gamma \gamma \to \mu^-\bar \mu^+ } & = & 4 {z} (1-z) 
 \left (1 -\frac{M_\mu^2}{M_\perp^2} \right ){\cal{H}}^{\gamma \gamma }(x_1, x_2,\bm{q}_{\sT}^2)
\,,\\
{\cal C}^{\gamma \gamma \to \mu^-\mu^+} & = & -z(1-z) \bigg (1-\frac{M_\mu^2}{M_\perp^2} \bigg )^2 \,\bigg [2 {\cal{I}}^{\gamma \gamma }(x_1, x_2,\bm{q}_\sT^2) - {\cal{L}}^{\gamma \gamma }(x_1, x_2,\bm{q}_\sT^2)\bigg ]\,,
\label{eq:ABCgam2}
\end{eqnarray} 
where we have defined the function
\begin{eqnarray}
{\cal A}_{1}& =& 2 \left [ {z}^2 + (1-z)^2 + 4 z (1- z)\left (1-\frac{M_\mu^2}{M_\perp^2} \right )\frac{M_\mu^2}{M_\perp^2}\right ]\,\label{eq:A1gam} 
\end{eqnarray}
and the convolutions adopted are the ones in 
Eqs.~(\ref{eq:Fqq})-(\ref{eq:Ngg}), 
(\ref{eq:Hqq})-(\ref{eq:Hgg}), (\ref{eq:I})-(\ref{eq:L}), with the obvious substitutions $f_1^g \to f_1^\gamma$ and $h_1^{\perp \,g} \to 
h_1^{\perp\,\gamma}$. We note that, because of the Drell-Yan background 
process, the cleanest way to extract $h_1^{\perp\,\gamma}$ in hadronic 
collisions would be through the measurement of a
$\cos 4(\phi_\perp-\phi_\sT)$ asymmetry, or else a selection that suppresses $s$-channel muon pair production, like a sizable lower $Q^2$ cut, should be considered.

\section{\label{Colorflow}Factorization issues and process dependent color factors}

The results in this paper have assumed TMD factorization. As is well-known, initial and final state interactions generally lead to modifications of the expressions depending on the process under consideration. Already at the level of resumming the corresponding collinear gluons  into the gauge links required for color gauge invariance, problems can arise with factorization~\cite{Rogers:2010dm}. Such factorization breaking effects show up in the dijet and heavy quark pair production cases, considered in the previous section. Despite these problems with TMD factorization for the differential (unintegrated) cross sections, transverse momentum weighted expressions, for $h_1^{\perp g}$ defined as
\begin{equation}
h_1^{\perp g (2) [U]}(x) \equiv \int d^2 \bm{p}_\sT 
\left(\bm{p}_\sT^2/2 M^2\right)^2 h_1^{\perp g [U]}(x, \bm{p}_\sT^2),
\end{equation}
{\em can} be factorized, but they appear with specific factors for different diagrams in the partonic subprocess~\cite{Bomhof:2006dp,Bomhof:2006ra}. This is simplest in cases where only the transverse momenta in just one of the hadrons matter~\cite{Buffing:2011mj}. The various factors result from the initial and final state interactions that can contribute differently in different subprocesses. By studying all weightings one can calculate and quantify the process dependence and the nonuniversality of the TMDs involved. Subsequently, one can then re-collect these transverse moments and express any gauge link dependent TMD into a finite number of TMDs of definite rank, e.g.\ three different `pretzelocity' functions ($h_{1T}^\perp$) in the case of quark TMDs~\cite{Buffing:2012sz}. Each of the functions corresponds to a Fourier transform of a well-defined operator combination in the defining matrix element.

Also, when writing down TMD factorized expressions for the processes $ep \to e' Q \bar{Q} X$, $pp \to \gamma \gamma X$ or $p p \to H/\eta_c/\chi_{c0}/... X$ that have been suggested as clean and safe ways to extract $h_1^{\perp g}(x,\bm p_\sT^2)$, one needs to be aware that one is not extracting a single TMD function, but a combination of several functions. For example, the $\gamma^* g \to Q \bar{Q}$ subprocess that transports a color octet initial state into a color octet final state, will lead to a gluon correlator with a different gauge link structure as compared to the subprocess where two gluons fuse to produce a color singlet final state. 

Using transverse weightings for the case of $h_1^{\perp g}$, the gauge link dependent TMDs can be expressed in a set of five universal TMDs~\cite{Buffing:2013notpublishedyet},
\begin{eqnarray}
h_1^{\perp g [U]}(x,\bm p_\sT^2) & = & h_1^{\perp g (A)}(x,\bm p_\sT^2) 
+ \sum_{c=1}^4 C_{GG,Bc}^{[U]} \, h_1^{\perp g (Bc)}(x,\bm p_\sT^2),
\label{eq:h1perp}
\end{eqnarray}
all of which have the same azimuthal dependence. Four of them, labeled (Bc), are gluonic pole matrix elements with in this case two soft gluonic pole contributions (and hence $T$-even), coming with a link dependent factor. There are multiple functions because the color trace can be performed in different ways. The function labeled with (A) does not contain a gluonic pole contribution (hence also $T$-even) and it contributes with factor unity in all situations. For further details on the definition of these functions and the relevant (calculable) gluonic pole factors we refer to Ref.~\cite{Buffing:2013notpublishedyet}.

As mentioned earlier it depends on the process under consideration which of the color structures appear. In $ep \to e' Q \bar{Q} X$ and in all the processes with a colorless final state, $pp \to \gamma \gamma X$ and $p p \to H/\eta_c/\chi_{c0}/... X$, only the two functions $h_1^{\perp g (A)}$ and $h_1^{\perp g (B1)}$ appear in the combination $h_1^{\perp g [gg\rightarrow \text{color singlet}]}=h_1^{\perp g (A)}+h_1^{\perp g (B1)}$, despite the different gauge link 
structures.
For $pp \to Q \bar{Q} X$ also the other functions appear due to the more complicated color flow of the diagram(s) involved. For example, in the case of $gg\rightarrow q\bar{q}$ in the hard scattering amplitude, there are multiple Feynman diagrams contributing to the process and all five functions in Eq.~(\ref{eq:h1perp}) are required.
Even if the basic tree level values of the gluonic pole coefficients $C_{GG,c}^{[U]}$ (with $c=1,\ldots,4$) can be calculated straightforwardly, one must be careful in those cases in which transverse momenta of more than just one hadron are involved, since these hadron-hadron scattering processes do not factorize in general. Therefore the relative strengths of the various azimuthal dependences attributed to linearly polarized gluons need further study.

\section{Summary and conclusions}

In this paper we have presented expressions for azimuthal asymmetries that arise in heavy quark and muon pair production due to the fact that gluons and also photons inside unpolarized hadrons can be linearly polarized. We studied these asymmetries for both electron-hadron and hadron-hadron scattering, not taking into account the presence of initial and final state interactions, which however modify the expressions by $N_c$-dependent pre-factors if not hampering TMD factorization altogether. For the processes considered in this paper this was addressed at the end in Sect.\ \ref{Colorflow}. 

First we considered the case of heavy quark pair production in electron-hadron scattering: $ep \to e' Q \bar{Q} X$. We calculated the maximal asymmetries ($R$ and $R'$) for two specific angular dependences. These turn out to be very sizable in certain transverse momentum regions. This finding, together with the relative simplicity of the measurements, are very promising concerning a future extraction of the linearly polarized gluon distribution $h_1^{\perp \; g}$ at EIC or LHeC. A similar conclusion applies to the linearly polarized photon distribution inside unpolarized protons through muon pair production. These measurements can be made relatively free from background, where for heavy quark pair production the contributions from intrinsic charm and bottom can be suppressed by restricting to the $x$ region below 0.1 (of course, the study of the polarization of intrinsic heavy quarks is of interest in itself) and for muon pair production the Drell-Yan background can be cut out by kinematic constraints. For the case of dijet production the asymmetries are expected to be smaller and background subtractions may be more involved. 

Next we considered heavy quark and muon pair production in hadron-hadron collisions. In this case the main concern is the breaking of factorization due to ISI and FSI. As explained in Sect.\ \ref{Colorflow}, cross sections can be expressed in terms of five universal $h_1^{\perp \; g}$ TMDs, in process dependent combinations, if factorization holds to begin with. It turns out that the $ep \to e' Q \bar{Q} X$ process probes the same combination of two of the five universal functions as processes like $pp \to \gamma \gamma X$ or $p p \to H/\eta_c/\chi_{c0}/... X$. This restricted universality can be tested experimentally, 
using RHIC or LHC data. In the process $pp \to Q \bar{Q} X$ factorization is expected to be broken, therefore, it is of interest to compare the extractions of 
$h_1^{\perp \; g}$ from 
$ep \to e' Q \bar{Q} X$ and $pp \to Q \bar{Q} X$, in order to learn about the size and importance of the factorization breaking effects. 
A further comparison to $ep \to e' \mu^-\mu^+X$ and $pp \to \mu^-\mu^+ X$ will be very interesting in this respect too, since these processes should not 
suffer from factorization breaking effects due to ISI/FSI. It will also teach us about the linearly polarization of photons in unpolarized protons. A further comparison to the distribution of linearly polarized photons `inside' electrons could also be very instructive. 
In this respect any high energy $e^+ e^-$, $e p$ and $pp$ scattering experiment can contribute valuably to such interesting comparisons.

\begin{acknowledgments}
This research is part of the research program
of the ``Stichting voor Fundamenteel Onderzoek der Materie (FOM)", which is financially 
supported by the ``Nederlandse Organisatie voor Wetenschappelijk Onderzoek (NWO)". We acknow\-ledge financial support from the European Community under the FP7 ``Capacities - Research Infrastructures'' program (HadronPhysics3 and the Grant Agreement 283286) and the ``Ideas'' program QWORK (contract 320389). C.P.\ would like to thank the Department of Physics of the University of Cagliari, and 
INFN, Sezione di Cagliari, where part of this work was performed.

\end{acknowledgments}


\begin{thebibliography}{99}

\bibitem{Mulders:2000sh}
  P.~J.~Mulders and J.~Rodrigues,
  Phys.\ Rev.\  D {\bf 63}, 094021 (2001).

\bibitem{Meissner:2007rx}
  S.~Meissner, A.~Metz, and K.~Goeke,
  Phys.\ Rev.\ {\bf D76}, 034002 (2007).

\bibitem{Boer:2009nc}
  D.~Boer, P.~J.~Mulders, and C.~Pisano,
  Phys.\ Rev.\ D. {\bf 80},  094017(2009).

\bibitem{Rogers:2010dm}
  T.~C.~Rogers and P.~J.~Mulders,
  Phys.\ Rev.\  D {\bf 81}, 094006 (2010).

\bibitem{Boer:2010zf}
  D.~Boer, S.~J.~Brodsky, P.~J.~Mulders, and C.~Pisano,
  Phys.\ Rev.\ Lett.\  {\bf 106}, 132001 (2011).

\bibitem{Qiu:2011ai} 
  J.~-W.~Qiu, M.~Schlegel, and W.~Vogelsang,
  Phys.\ Rev.\ Lett.\  {\bf 107}, 062001 (2011).

\bibitem{Metz:2011wb} 
  A.~Metz and J.~Zhou,
  Phys.\ Rev.\ D {\bf 84}, 051503 (2011).

\bibitem{Dominguez:2011br} 
  F.~Dominguez, J.~-W.~Qiu, B.~-W.~Xiao, and F.~Yuan,
  Phys.\ Rev.\ D {\bf 85}, 045003 (2012).

\bibitem{Schafer:2012yx} 
  A.~Sch\"afer and J.~Zhou,
  Phys.\ Rev.\ D {\bf 85}, 114004 (2012).

\bibitem{Akcakaya:2012si} 
  E.~Akcakaya, A.~Sch\"afer, and J.~Zhou,
  Phys.\ Rev.\ D {\bf 87}, 054010 (2013).

\bibitem{Nadolsky:2007ba} 
  P.~M.~Nadolsky, C.~Balazs, E.~L.~Berger, and C.~-P.~Yuan,
  Phys.\ Rev.\ D {\bf 76}, 013008 (2007).

\bibitem{Catani:2010pd} 
  S.~Catani and M.~Grazzini,
  Nucl.\ Phys.\ B {\bf 845}, 297 (2011).

\bibitem{deFlorian:2012mx} 
  D.~de Florian, G.~Ferrera, M.~Grazzini and D.~Tommasini,
  JHEP {\bf 1206}, 132 (2012).

\bibitem{Sun:2011iw} 
  P.~Sun, B.~-W.~Xiao, and F.~Yuan,
  Phys.\ Rev.\ D {\bf 84}, 094005 (2011).

\bibitem{Boer:2013fca} 
  D.~Boer, W.~J.~den~Dunnen, C.~Pisano, and M.~Schlegel,
  arXiv:1304.2654 [hep-ph].

\bibitem{Boer:2011kf} 
  D.~Boer, W.~J.~den~Dunnen, C.~Pisano, M.~Schlegel, and W.~Vogelsang,
  Phys.\ Rev.\ Lett.\  {\bf 108}, 032002 (2012).

\bibitem{Boer:2012bt} 
  D.~Boer and C.~Pisano,
  Phys.\ Rev.\ D {\bf 86}, 094007 (2012).

\bibitem{Grzadkowski:1992sa} 
  B.~Grzadkowski and J.~F.~Gunion,
  Phys.\ Lett.\ B {\bf 294}, 361 (1992).

\bibitem{Gunion:1994wy} 
  J.~F.~Gunion and J.~G.~Kelly,
  Phys.\ Lett.\ B {\bf 333}, 110 (1994).

\bibitem{Kramer:1993jn} 
  M.~Kr\"amer, J.~H.~K\"uhn, M.~L.~Stong, and P.~M.~Zerwas,
  Z.\ Phys.\ C {\bf 64}, 21 (1994).

\bibitem{Gounaris:1997ef} 
  G.~J.~Gounaris and G.~P.~Tsirigoti,
  Phys.\ Rev.\ D {\bf 56}, 3030 (1997)
  [Erratum-ibid.\ D {\bf 58}, 059901 (1998)].

\bibitem{Asner:2001ia} 
  D.~M.~Asner, J.~B.~Gronberg, and J.~F.~Gunion,
  Phys.\ Rev.\ D {\bf 67}, 035009 (2003).

\bibitem{Kamal:1995ct} 
  B.~Kamal, Z.~Merebashvili, and A.~P.~Contogouris,
  Phys.\ Rev.\ D {\bf 51}, 4808 (1995)
  [Erratum-ibid.\ D {\bf 55}, 3229 (1997)].

\bibitem{Jikia:1996bi} 
  G.~Jikia and A.~Tkabladze,
  Phys.\ Rev.\ D {\bf 54}, 2030 (1996).

\bibitem{Melles:1998gu} 
  M.~Melles and W.~J.~Stirling,
  Phys.\ Rev.\ D {\bf 59}, 094009 (1999).

\bibitem{Jikia:2000rk} 
  G.~Jikia and A.~Tkabladze,
  Phys.\ Rev.\ D {\bf 63}, 074502 (2001).

\bibitem{Kniehl:2009kh} 
  B.~A.~Kniehl, A.~V.~Kotikov, Z.~V.~Merebashvili, and O.~L.~Veretin,
  Phys.\ Rev.\ D {\bf 79}, 114032 (2009).

\bibitem{Buffing:2011mj} 
  M.~G.~A.~Buffing and P.~J.~Mulders,
  JHEP {\bf 1107}, 065 (2011).

\bibitem{Buffing:2012sz} 
  M.~G.~A.~Buffing, A.~Mukherjee, and P.~J.~Mulders,
  Phys.\ Rev.\ D {\bf 86}, 074030 (2012).

\bibitem{Buffing:2013notpublishedyet}
  M.~G.~A.~Buffing, A.~Mukherjee, and P.~J.~Mulders,
     arXiv:1306.5897 [hep-ph].

\bibitem{Boer:1997nt}
  D.~Boer and P.~J.~Mulders,
  Phys.\ Rev.\  D {\bf 57}, 5780 (1998).

\bibitem{Boer:1999mm} 
  D.~Boer,
  Phys.\ Rev.\ D {\bf 60}, 014012 (1999).
  
\bibitem{Boer:2002ju} 
  D.~Boer, S.~J.~Brodsky and D.~S.~Hwang,
  Phys.\ Rev.\ D {\bf 67}, 054003 (2003)
  [hep-ph/0211110].

\bibitem{Bacchetta:2006tn}
  A.~Bacchetta, M.~Diehl, K.~Goeke, A.~Metz, P.~J.~Mulders, and M.~Schlegel,
  JHEP {\bf 0702}, 093 (2007).

\bibitem{Boer:2007nd}
  D.~Boer, P.~J.~Mulders, and C.~Pisano,
  Phys.\ Lett.\  B {\bf 660}, 360 (2008).

\bibitem{Mirkes:1997ru} 
  E.~Mirkes and S.~Willfahrt,
  Phys.\ Lett.\ B {\bf 414}, 205 (1997).


\bibitem{Brodkorb:1994de} 
  T.~Brodkorb and E.~Mirkes,
  Z.\ Phys.\ C {\bf 66}, 141 (1995).

\bibitem{Brodsky:1980pb}
  S.~J.~Brodsky, P.~Hoyer, C.~Peterson, and N.~Sakai,
  Phys.\ Lett.\ B {\bf 93}, 451 (1980).

\bibitem{Brodsky:1981se}
  S.~J.~Brodsky, C.~Peterson, and N.~Sakai,
  Phys.\ Rev.\ D {\bf 23}, 2745 (1981).

\bibitem{Pumplin:2005yf}
  J.~Pumplin,
  Phys.\ Rev.\ D {\bf 73}, 114015 (2006).

\bibitem{Pumplin:2007wg}
  J.~Pumplin, H.~L.~Lai, and W.~K.~Tung,
  Phys.\ Rev.\ D {\bf 75}, 054029 (2007).

\bibitem{Chang:2011vx}
  W.~-C.~Chang, and J.~-C.~Peng,
  Phys.\ Rev.\ Lett.\  {\bf 106}, 252002 (2011).

\bibitem{Chang:2011du}
  W.~-C.~Chang and J.~-C.~Peng,
  Phys.\ Lett.\ B {\bf 704}, 197 (2011).

\bibitem{Mirkes:1997uv} 
  E.~Mirkes,
  Habilitation thesis, Universit\"at Karlsruhe,
  hep-ph/9711224.


\bibitem{Bjorken:1966kh}
  J.~D.~Bjorken and M.~C.~Chen,
  Phys.\ Rev.\  {\bf 154}, 1335 (1967) [Erratum-ibid.\ {\bf 162},  
  1750 (1967)].

\bibitem{Brodsky:1966vh}
  S.~J.~Brodsky and S.~C.~C.~Ting,
  Phys.\ Rev.\  {\bf 145}, 1018 (1966).

\bibitem{Tannenbaum:1968zz}
  M.~J.~Tannenbaum,
  Phys.\ Rev.\  {\bf 167}, 1308 (1968).

\bibitem{Gluck:2002fi} 
  M.~Gl\"uck, C.~Pisano, and E.~Reya,
  Phys.\ Lett.\ B {\bf 540}, 75 (2002).

\bibitem{Gluck:2002cm}
  M.~Gl\"uck, C.~Pisano, E.~Reya, and I.~Schienbein,
  Eur.\ Phys.\ J.\  C {\bf 27}, 427 (2003).
%
\bibitem{Kniehl:2004fy}
  B.~A.~Kniehl, G.~Kramer, I.~Schienbein, and H.~Spiesberger,
  Phys.\ Rev.\  D {\bf 71}, 014018 (2005).

\bibitem{Anselmino:2004nk}
  M.~Anselmino {\it et al.}, 
  Phys.\ Rev.\  D {\bf 70}, 074025 (2004).

\bibitem{Gluck:1977zm}
  M.~Gl\"uck, J.~F.~Owens, and E.~Reya,
  Phys.\ Rev.\  D {\bf 17}, 2324 (1978).

  \bibitem{Brodsky:1997jk}   
  S.~J.~Brodsky and P.~Huet,
  Phys.\ Lett.\ B {\bf 417}, 145 (1998).

  
\bibitem{Bomhof:2006dp}
  C.~J.~Bomhof, P.~J.~Mulders, and F.~Pijlman,
  Eur.\ Phys.\ J.\  C {\bf 47}, 147 (2006).

\bibitem{Bomhof:2006ra} 
  C.~J.~Bomhof and P.~J.~Mulders,
  JHEP {\bf 0702}, 029 (2007).


\bibitem{Dominguez:2010xd}
     F.~Dominguez, B~-W.~Xiao, and F.~Yuan,
     Phys.\ Rev.\ Lett.\  {\bf 106}, 022301 (2011).


\end{thebibliography}
\end{document}